\RequirePackage{fix-cm}
\RequirePackage{amsmath}
\documentclass[smallextended]{svjour3}       % onecolumn (second format)
\smartqed  % flush right qed marks, e.g. at end of proof
\usepackage{graphicx}
\usepackage{mathptmx}      % use Times fonts if available on your TeX system
\usepackage{newtxtext,newtxmath}
\usepackage{color}
\usepackage{xcolor}
\usepackage{empheq}
\usepackage{nicefrac}
\usepackage{tikz}
\usepackage{enumitem}
\usetikzlibrary{shapes,arrows}

\definecolor{dunkelgruen}{rgb}{0,0.65,0}
\definecolor{darkblue}{RGB}{0,50,100}

\usepackage[colorlinks=true, linkcolor=darkblue, 
            citecolor=darkblue, urlcolor=darkblue]{hyperref}
\journalname{Reviews of Modern Plasma Physics}
\newcommand{\fett}[1]{\boldsymbol{#1}}
\newcommand{\dd}{{\rm{d}}}
\newcommand{\ii}{{\rm{i}}}
\newcommand{\be}{\begin{equation}}
\newcommand{\ee}{\end{equation}}
\newcommand{\nab}{\fett{\nabla}}
\newcommand{\nabx}{\fett{\nabla}_{\!\fett{x}}}
\newcommand{\nabq}{\fett{\nabla}_{\fett{q}}}

\newcommand{\DOI}[2]{ 
 \href{http://dx.doi.org/#1}{\color{darkblue}#2\!}
}
\newcommand{\ADS}[2]{ 
 \href{https://ui.adsabs.harvard.edu/abs/#1}{\color{darkblue}#2\!}
}
\newcommand{\arXiv}[1]{ 
 \href{https://arxiv.org/abs/#1}{\color{darkblue}arXiv:#1\!}
}
\newcommand*\widefbox[1]{\fbox{\hspace{1em}#1\hspace{1em}}}

\newcommand*\circled[1]{\tikz[baseline=(char.base)]{
            \node[shape=circle,draw,inner sep=2pt] (char) {#1};}}
\newcommand{\xistar}{\fett{\xi}_{\!\!\star}}
\newcommand{\dotxistar}{\dot{\fett{\xi}}_{\!\!\star}}

\newcommand{\apj}{{Astrophys. J.} }
\newcommand{\apjl}{{Astrophys. J. Lett.} }
\newcommand{\apjs}{{Astrophys. J. Suppl.} }

\newcommand{\mnras}{{Mon. Not. Roy. Astron. Soc.} }
\newcommand{\mnrasl}{{Mon. Not. Roy. Astron. Soc.} {\it Letters} }

\newcommand{\physrep}{   {Phys. Reports} }

\newcommand{\jcap}{ {J. Cosmology Astropart. Phys.} }
\newcommand{\jhep}{ {J. High Energy Phys.} }

\definecolor{lime}{HTML}{A6CE39}
\DeclareRobustCommand{\orcidicon}{
	\begin{tikzpicture}
	\draw[lime, fill=lime] (0,0) 
	circle [radius=0.14] 
	node[white] {{\fontfamily{qag}\selectfont \tiny ID}};
	\draw[white, fill=white] (-0.0625,0.095) 
	circle [radius=0.007];
	\end{tikzpicture}
	\hspace{-2mm}
}

\foreach \x in {A, ..., Z}{%
	\expandafter\xdef\csname orcid\x\endcsname{\noexpand\href{https://orcid.org/\csname orcidauthor\x\endcsname}{\noexpand\orcidicon}}
}

\begin{document}

\title{Cosmological Vlasov--Poisson equations for dark matter} 

\subtitle{Recent developments and connections to selected plasma problems}

\titlerunning{Cosmological Vlasov--Poisson equations}        

\author{Cornelius Rampf}

\institute{C.\,Rampf\orcidA{}\at
  Department of Mathematics, University of Vienna,  1090 Vienna, Austria \\
  Department of Astrophysics, University of Vienna, 1180 Vienna, Austria \\
  \email{cornelius.rampf@univie.ac.at} 
}

\date{Received: date / Accepted: date}

\maketitle

\begin{abstract}
The cosmic large-scale structures of the Universe are mainly the result of the gravitational instability of initially small density fluctuations in the dark-matter distribution. Dark matter appears to be initially cold and behaves as a continuous and collisionless medium on cosmological scales, with evolution governed by the gravitational Vlasov--Poisson equations.  Cold dark matter can accumulate very efficiently at focused locations, leading to a highly non-linear filamentary network with extreme matter densities. Traditionally, investigating the non-linear Vlasov--Poisson equations was typically reserved for massively parallelised numerical simulations. Recently, theoretical progress has allowed us to analyse the mathematical structure of the first infinite densities in the dark-matter distribution by elementary means. We review related advances, as well as provide intriguing connections to classical plasma problems, such as the beam-plasma instability.

% list of PACS: https://ufn.ru/en/pacs/
% %$98.80.−k$  \and 95.35.+d \and 52.65.Ff \and 52.65.Vv \and 02.40.Xx}
\end{abstract}

\maketitle

\tableofcontents

\section{Basic problem}\label{sec:VPintro}

\begin{figure}
  \begin{center}

   \includegraphics[width=0.95\textwidth]{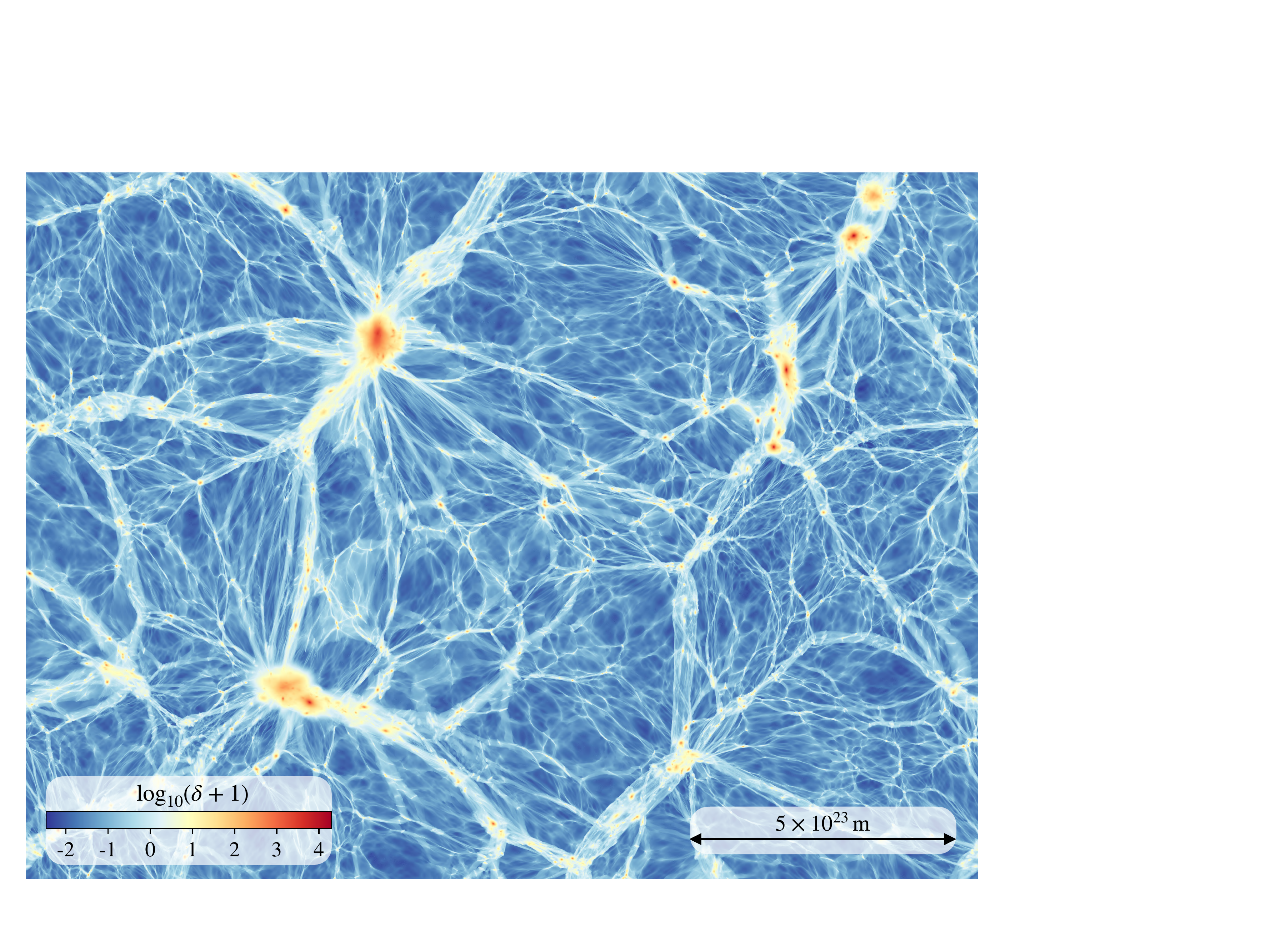}

  \end{center}
\caption{Simulation result showing the logarithm of the over-density $\delta+1 = \rho/\bar \rho$ of the dark-matter distribution on cosmic scales (thickness of slice is $l \sim 10^{21}$m); figure adapted from \cite{Stucker:2017nmi}. Here, $\rho$ is the total density while~$\bar \rho(t)$ is an isotropic dilution factor stemming from the volume expansion of the Universe. }
\label{fig:LSS}
%https://wwwmpa.mpa-garching.mpg.de/paper/singlestream2017/density.html
\end{figure}

What is the nature of dark matter and dark energy, which together make up about 95\% of the Universe’s energy content? One way to address such a question is to investigate the evolution of cosmic structures on the largest observable length scales~(Fig.\,\ref{fig:LSS}).

Dark matter constitutes the bulk part of the overall matter distribution. Apart from gravitational interactions, dark matter appears to be extremely weakly interacting (e.g.\ \cite{Aghanim:2018eyx,Arcadi:2017kky,Smorra:2019qfx,Kunz:2016yqy}), thereby justifying the validity of the collisionless limit on cosmological scales ($\sim10^{22}\,{\rm m}-10^{26}$\,m).  The gravitational evolution of such a collisionless medium is governed by the cosmic Vlasov--Poisson equations, which describe how the dark-matter distribution $f=f(\fett{x},\fett{p},t)$ evolves in the six-dimensional phase-space, 
\be \label{eq:VP}
  \frac{\dd f}{\dd t} = \frac{\partial f}{\partial t} + \frac{\fett{p}}{m a^2} \cdot \nabx f 
    - m (\nabx \phi) \cdot \fett{\nabla}_{\fett{p}} f = 0\,,
    \quad  \qquad  \nabla^2_{\fett{x}} \phi = 4\pi G \bar \rho(t)\, a^2 \delta(\fett{x},t) \,,
\ee
where $\delta$ is the matter density contrast defined through the matter density
\be 
  \rho = \bar \rho(t) [1+ \delta(\fett{x},t) ]\, , \qquad \quad  \rho = \int f \!(\fett{x},\fett{p},t) \,\,\dd^3 p \,.
\ee
Here, $a=a(t)$ is the cosmic scale factor that encodes the mean expansion of the Universe, and is subject to the 
Friedmann equation $(\dot a/a)^2 = 8\pi G \bar \rho/3 - K c^2/a^2 + \Lambda c^2/3$, where $K$ is a uniform curvature of space and $\Lambda$ is a cosmological constant associated with dark energy, while $\bar \rho \sim a^{-3}$ is the mean matter density which is diluting in time, due to the so-called Hubble expansion of the spatial volume of the Universe.
The Friedmann equation determines the functional relationship between $a$ and $t$ for given matter/energy content, and pins down the speed of the Hubble expansion. For example, for the Einstein--de Sitter  Universe \cite{EdSuniverse,Bernardeau:2001qr}, which is a simplified cosmological model with  $K=0=\Lambda$ and a common starting point for theoretical derivations, we have $a\propto t^{2/3}$. For the  observationally preferred ``$\Lambda$ cold dark matter'' ($\Lambda$CDM) model \cite{Peebles:1994xt,WMAP:2012nax} which is spatially flat ($K=0$) and has a cosmological constant of $\Lambda \simeq 1.1 \times 10^{-52} {\rm m}^{-2}$ in Planck units~\cite{Aghanim:2018eyx}, we have $a \propto \Lambda^{-1/3} \sinh^{2/3}(3\sqrt{\Lambda}c t/2) \propto t^{2/3} +O(t^{8/3})$.

The  distribution function~$f$ depends on the ``co-moving'' variables $\fett{x}$ and $\fett{p}$ (co-moving with the Hubble expansion), which are linked through a canonical transformation to the physical position $\fett{X}$ and physical momentum $\fett{P}$ with 
\be \label{eqs:canonicaltrafo}
  \fett{x} = \fett{X}/a\, , \qquad \qquad   \fett{p}= a \fett{P} - m \, \dot a\, \fett{X}\,.
\ee
Note that the co-moving momentum $\fett{p}$ has a trivial %and homogeneous 
shift $\propto  \dot a\, \fett{X}$ subtracted out, due to the said expansion of the Universe.  
Consequently, the distribution function that solves the Vlasov--Poisson equations~\eqref{eq:VP} tracks only the non-trivial aspects of the dark-matter phase-space, while the position~$\fett{X}$ and momentum~$\fett{P}$ in the physical phase-space can be easily recovered through~\eqref{eqs:canonicaltrafo} in the Newtonian limit \cite{Buchert:1995fz}.
The use of co-moving variables is ubiquitous in %theoretical and numerical 
cosmology, and in fact comprises one of the key ingredients for establishing mathematical analyticity in the dark-matter phase-space for sufficiently short times; see section~\ref{sec:mastereqs} for details.

Finally, note that the Vlasov--Poisson equations~\eqref{eq:VP} imply a Hamiltonian formulation, 
with corresponding one-particle Hamiltonian 
\be
  H( \fett{x}, \fett{p},t) = \frac{|\fett{p}|^2}{2ma^2} + m \phi(\fett{x},t) \,,
\ee
and equations of motion
\be \label{eqs:HamiltonEoMs}
  \dot{\fett{x}} = \fett{\nabla}_{\!\fett{p}} H = \frac{1}{ma^2} \fett{p} , \qquad \quad \dot{\fett{p}} = - \nabx H = - m \nabx \phi .
\ee
Thanks to the Hamiltonian nature,  the system is constrained by Liouville's theorem, implying that the dark-matter phase-space distribution is incompressible and thus devoid of any (severe) disruptions; we will come back to this in section~\ref{sec:PSC}.

%%%%%%%%%%%%%%%%%%%%%%%%%%%%%%%%%%%
\section{Topology of the dark-matter phase-space and shell-crossings}

\subsection{The dark-matter sheet and suitable parametrisation}

Cosmological observations indicate that the initial dark-matter distribution is extremely cold (e.g.~\cite{Aghanim:2018eyx,Kunz:2016yqy,Thomas:2016iav,Gilman:2019nap,Ilic:2020onu}). Furthermore, any residual weak thermalisation of the initial dark-matter distribution, if present, should have negligible impact on the gravitational dynamics on the considered macroscopic scales. Therefore, it is customary in cosmology to assume the limiting case of initially perfect coldness, which is also assumed in the present paper.
A perfectly cold distribution has vanishing (thermal) velocity dispersion, implying that at sufficiently early times, the whole kinetic information is encoded into a single velocity field $\fett{v} := \fett{p} /(m a^2 \partial_t a)$. Nevertheless, non-zero velocity dispersion (and thus an effective non-zero temperature) is generated during the later stages of the gravitational dynamics, which is discussed next.

\setcounter{footnote}{-1}

\begin{figure}
 \begin{center}
  \includegraphics[width=0.97\textwidth]{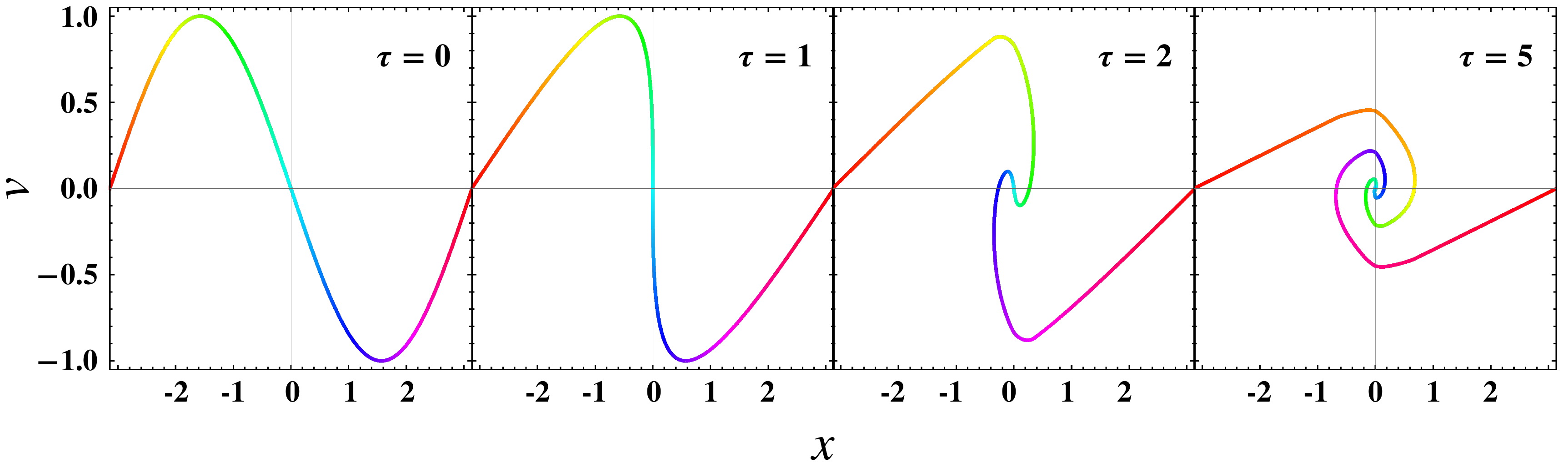} 
 \end{center}
\caption[caption]{Evolution of the dark-matter sheet in the phase-space, employing a simplified one-dimensional setup on the torus with $x \in [ -\pi, \pi )$ in arbitrary units in a (fictitious) Universe filled only with dark matter. For convenience we have added individual colour markings to the dark-matter fluid particles (i.e., the Lagrangian coordinate), thereby highlighting the fluid motion during the temporal evolution. Shown results were obtained using a particle in cell simulation with 8192~fluid particles\protect\footnotemark, initialised at $\tau=0$ with a sine-wave velocity profile and vanishing density contrast~\cite{Rampf:2019nvl} ($\tau\propto t^{2/3}$ where $t$ is standard time). The second panel shows the critical instant when the dark-matter sheet folds for the first time (here at $\tau=1$ and $x=0$), which is usually denoted with shell-crossing or stream crossing. The third and fourth panels show multi-streaming regions where multiple matter streams occupy the same current position.}
\label{fig:phasespace-color}
\end{figure}

The topology of the phase-space distribution is such that it is confined on an infinitesimally thin sheet along the momentum direction and thus occupies, at all times, only  a three-dimensional hypersurface in the full 6D phase-space (the Lagrangian sub-manifold). The resulting dark-matter sheet is illustrated in Fig.\,\ref{fig:phasespace-color} in a simplified set-up in 1+1 dimensions.
During the course of the gravitational evolution, the dark-matter sheet never tears and remains connected, due to the Hamiltonian nature of the system. 
Nevertheless, gravitational interactions can generate folds in the dark-matter sheet.
At those folds, the number of matter streams that occupy the same current position changes. 
An example of this is shown in Fig.\,\ref{fig:phasespace-color}, which indicates the emergence of a proliferation of multiple fluid streams.

\footnotetext{\label{footnote1Dsim}Available from \url{https://bitbucket.org/ohahn/cosmo_sim_1d}.}

Obviously, solving the Vlasov--Poisson equations~\eqref{eq:VP} in ``full 6D'' is vastly excessive, since the dark-matter distribution is confined on a three-dimensional hypersurface. Therefore, it is customary, in both theory and numerics, to employ an efficient parametrisation of the dark-matter sheet, that essentially tracks how the positions and momenta of dark-matter particles change on the three-dimensional hypersurface. Precisely such a ``map'' is realised when introducing Lagrangian coordinates, which in cosmology are usually denoted with $\fett{q}$ and can be thought of as a continuous set of particle labels (denoted by the colour marking of the dark matter sheet in Fig.\,\ref{fig:phasespace-color}).

Indeed, introducing the Lagrangian map $\fett{q} \mapsto \fett{x}(\fett{q},\tau)$
from initial ($\tau =0$) position~$\fett{q}$ to the current Eulerian position $\fett{x}$ at a refined temporal variable $\tau$ ($\propto t^{2/3}$; see below), the Vlasov--Poisson equations for dark matter take the form (see e.g.~\cite{Bernardeau:2001qr,Matsubara:2007wj,Uhlemann:2018gzz})
\begin{empheq}[box=\widefbox]{align}
\label{eq:VPLag}
  \!\!\!\!\!\ddot {\fett{x}}(\fett{q},\tau) =   - \frac{3}{2\tau} \big[ \dot {\fett{x}}(\fett{q},\tau) +  \nabx \varphi(\fett{x}(\fett{q},\tau)) \big] \,, \qquad    \nabla_x^2 \varphi(\fett{x}(\fett{q},\tau)) = \frac{\delta(\fett{x}(\fett{q},\tau))}{\tau} \,,\!\!\!\!
\end{empheq}
where, for convenience, we employ a rescaled Poisson potential $\varphi =  \phi/(4\pi G \bar \rho a^2 \tau)$. 
Here and in the following, over-dots denote convective (i.e., total) derivatives with respect to the dimensionless temporal variable $\tau = a$; using the cosmic scale factor (or more generally, the  growth function of linear matter density fluctuations) as the temporal variable is another ingredient for establishing time-analytic solutions for sufficiently short times; see section~\ref{sec:mastereqs} for details. The Lagrangian map is defined such that $\fett{v} =: \dot{\fett{x}}(\fett{q},\tau)$ is the dark-matter velocity field (in units of lengths since $\tau=a$ is dimensionless), 
where $\fett{v}$ 
relates to the momentum (non-canonically) according to $\fett{p} = \fett{v} m a^2 \partial_t a$.
We note that equations~\eqref{eq:VPLag} are derived for the case when the  
Universe is spatially flat and solely composed of dark matter (i.e., the simplified Einstein--de Sitter model), but this could be easily generalised to the more realistic $\Lambda$CDM model if needed (see e.g.\ \cite{Rampf:2015mza,Matsubara:2015ipa,Rampf:2020ety}).  Nonetheless, from here on we will mostly work with this simplified case of an Einstein--de Sitter model to avoid unnecessary cluttering of the expressions.

The matter density contrast appearing in the equations~\eqref{eq:VPLag} 
can be exactly expressed through a mass conservation law using the Dirac-delta $\delta_{\rm D}^{(3)}$  \cite{Matsubara:2007wj,Taylor:1996ne,McDonald:2017ths,Morrison:2020}, here for the case when $\delta \to 0$ initially,
\be \label{eq:mass}
 \delta(\fett{x}(\fett{q},\tau)) +1  = \int \delta_{\rm D}^{(3)}\left[ \fett{x}(\fett{q},\tau)- \fett{x}(\fett{q}',\tau) \right] \, \dd^3 q' \,,
\ee 
which, roughly speaking, may be interpreted as the continuum limit $N \to \infty$ of a density contrast $\delta_N$ induced through a (hypothetical) collection of $N$ discrete particles with $\delta_{N}(\fett{x},\tau) +1 \sim \sum_{i=1}^N \delta_{\rm D}^{(3)}(\fett{x} - \fett{x}_i(\tau))$, where $\fett{x}_i(\tau)$ denotes the current position of the $i$th dark-matter particle (see e.g.~\cite{Buchert:2005xj} for a more precise argument).
Equation~\eqref{eq:mass} is the result of linking the law of mass conservation $\bar \rho (1+\delta(\fett{x},\tau)) \,\dd^3 x = \bar \rho \, \dd^3 q$ to the determinant of the Lagrangian mapping. 
Specifically, employing the composition property of the Dirac delta, the mass conservation~\eqref{eq:mass} can be equivalently written as
\be \label{eq:mass2}
  \delta(\fett{x},\tau) +1 =  \sum_{n ~\rm roots} \frac{1}{\left|\det [\nabq \fett{x}(\fett{q},\tau)]\right|_{\fett{q}= \fett{q}_n}} \,,
\ee
where $\nabq \fett{x} = \nabq \otimes \fett{x}^{\rm T}$ is the direct (dyadic) product, while $\fett{q}_n$ denotes the $n$th root of the equation $\fett{x}-\fett{x}(\fett{q},\tau) =\fett{0}$.
Intuitively, the sum in~\eqref{eq:mass2} appears since $n$ fluid streams occupy the same current position $\fett{x}$, and all fluid streams  obviously contribute to the density (see also the discussion further below and Fig.\,\ref{fig:map-density}).
Note that as long as the fluid is single stream, there is only a single root at $\fett{q}_1 = \fett{q}$, implying that mass conservation simplifies to $\delta(\fett{x}(\fett{q},\tau)) +1 = 1/J$ in Lagrangian coordinates, where $J=\det[\nabq \fett{x}(\fett{q},\tau)]$ is the Jacobian.

The Vlasov--Poisson equations~\eqref{eq:VPLag} have the simple physical interpretation of being essentially a Newtonian equation of motion, with the addition of a ``drag'' term $\propto (1/\tau) \dot {\fett{x}}$, which stems from the present choice of co-moving coordinates (co-moving with respect to a fixed position on a spatial grid that follows the overall Hubble expansion of the Universe).
Note the structural similarity of equations~\eqref{eq:VPLag}--\eqref{eq:mass} as compared to the Vlasov--Poisson equations for the beam-plasma instability, a classical plasma problem. There, a beam of (discretised) charged particles moves in a background neutral plasma. We will come back to such structural similarities in section~\ref{sec:plasma}.

Note that the Vlasov--Poisson equations~\eqref{eq:VPLag} are invariant under the transformation 
\be \label{eq:Heck}
  \fett{x}(\fett{q},\tau) \to \fett{x}'(\fett{q},\tau) = \fett{x}(\fett{q},\tau) + \fett{n}(\tau)
\ee
for arbitrary time-dependent boosts~$\fett{n}(\tau)$, provided that the gravitational potential transforms as $\varphi'(\fett{x}') = \varphi(\fett{x}) - (\dot{\fett{n}} + [2\tau/3] \ddot {\fett{n}}) \cdot \fett{x} + \fett h(\tau)$, for arbitrary $\fett h(\tau)$. This non-Galilean invariance has been first analysed by Heckmann \& Sch\"ucking in 1955 \cite{HeckmannSchucking1955}, and recently drew renewed attention in cosmology (e.g., \cite{Rampf:2019nvl,Ehlers:1996wg,Kehagias:2013yd,Peloso:2013zw,Creminelli:2013mca}). 
This invariance is of particular relevance when determining critical phenomena in the dark-matter phase-space (see section~\ref{sec:PSC}, in particular panel d in Fig.\,\ref{fig:PSC}).

Finally, observe  that the Vlasov--Poisson equations~\eqref{eq:VPLag} maintain their regularity for $\tau \to 0$, provided one imposes the following boundary conditions on the initial conditions \cite{Brenier:2003xs,Zheligovsky:2013eca} at time $\tau =0$ (denoted by the superscript ``ini''):
\be \label{eq:slaving}
  \delta^{\rm ini} =0 \,, \qquad \quad \fett{v}^{\rm ini} = -\nab \varphi^{\rm ini} ,
\ee
where $\fett{v}^{\rm ini}= \dot{\fett{x}}(\fett{q},\tau=0)$. Thus, with these boundary conditions, only the initial gravitational potential $\varphi^{\rm ini}$ at $\tau=0$ needs to be prescribed; see e.g.~\cite{Michaux:2020yis} for explicit instructions how this can be achieved in practice.
Mathematically, initialising the Vlasov--Poisson system at time ``zero'' (a.k.a.~the big bang) using~\eqref{eq:slaving} implies that the inflationary physics has been reduced to an infinitely thin boundary layer~\cite{Rampf:2015mza}.
Also note that the second condition in~\eqref{eq:slaving} immediately implies that $\nabx \times \fett{v}^{\rm ini} = \fett{0}$ which, due to Helmholtz's theorem for the conservation of vorticity flux, translates to $\nabx \times \dot{\fett{x}} = \fett{0}$ at all times along each matter stream  (but note that effective vorticity is generated in Eulerian coordinates through multi-streaming~\cite{Uhlemann:2018gzz,Pueblas:2008uv,Cusin:2016zvu}).
In the following we will mostly focus on solutions using the conditions~\eqref{eq:slaving}, since those are the only one currently available that have mathematical proofs of analyticity until some finite time; see sections~\ref{sec:sc-anal} and~\ref{sec:sc-random} for details.

\subsection{Theoretical challenge of resolving folds of the dark-matter sheet: Shell-crossings}\label{sec:shell-crossing}

\begin{figure}
  \begin{center}
  
  \includegraphics[width=0.98\textwidth]{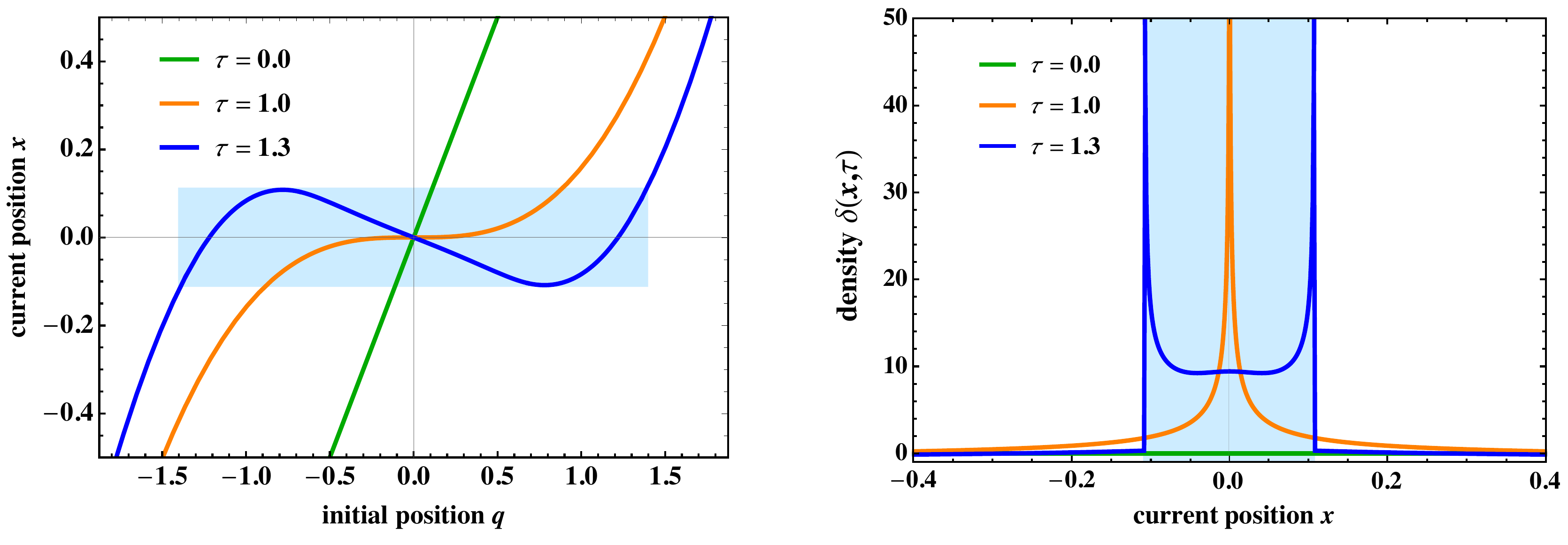}

  \end{center}
\caption{Shown is the Lagrangian map $q \mapsto x(q,\tau)$  from initial position $q$ to current position $x$ (left panel), and the density contrast $\delta= (\rho -\bar \rho)/\bar \rho$ (right panel), employing the analytical results of~\cite{Rampf:2019nvl} in the same simplified set-up as for Fig.\,\ref{fig:phasespace-color}. Note specifically that a torus geometry is used with $q, x \in [ -\pi, \pi )$, while shown is only the physically interesting regime.  Initially ($\tau=0$, green lines) the particle positions are perfectly uniformly distributed in space, which is reflected by a linear regression in the left panel,  and  by $\delta=0$ in the right panel resembling an initial homogeneous state. 
A critical instant in the dark-matter evolution arises when $\nabla_q x$ vanishes (corresponding to $\det[\nabq \fett{x}] =0$ in three space dimensions; cf.\ Eq.\,\ref{eq:mass2}), which appears in the present case at $\tau=1$ for the first time (orange line in left panel). This instant is usually called shell-crossing and is accompanied with an infinite density (center region in right plot). Furthermore, it marks the starting point when multiple streams of particles can occupy the same current position (blue shaded region shown at $\tau=1.3$), which comes with non-zero velocity dispersion.}
\label{fig:map-density}
\end{figure}

The instant when the dark-matter sheet begins to fold (central panel in Fig.\,\ref{fig:phasespace-color}), called the first shell-crossing, implies a critical moment in the phase-space evolution: at folds,  streams of dark-matter particles accumulate at focused locations leading to formally infinite densities. This can also be seen in Fig.\,\ref{fig:map-density} where we show, in the left panel, the current positions of matter streams (the particle trajectories for certain times) and, in the right panel, the corresponding matter density. Whenever the number of matter streams changes, the determinant of $\nabq \fett{x}$ vanishes, which, by virtue of equation~\eqref{eq:mass2} implies the said infinite density.

The critical instant, in time and space, when the dark-matter continuum changes from single stream to triple stream, is denoted for traditional reasons by  ``shell-crossing'' (since the first collapse models involved the study of ``matter shells'' in spherical symmetry).  
The first approximative theoretical model of the cosmic Vlasov--Poisson equation that predicts the appearance of the first shell-crossing traces back to the work of Ya.~B.~Zel'dovich in 1970 \cite{Zeldovich:1969sb}. Nowadays the approach is called Zel'dovich solution (in 1D) or  Zel'dovich approximation (beyond 1D), and, despite its age, comprises one of the most successful and insightful theoretical models in cosmic structure formation (e.g.~\cite{Buchert:1992ya,Valageas:2010rx,White:2014gfa,McQuinn:2015tva}). We review this model in section~\ref{sec:pertZA++}, but note that the Zel'dovich approximation constitutes an exact solution of the Vlasov--Poisson equations in a fictitious one-dimensional setup until the first shell-crossing \cite{Novikov:2010ta,ZentsovaChernin1980}.
Here and in the following, a ``one-dimensional setup'' can be established by providing 3D initial conditions that only depend on one space coordinate (i.e., a one-dimensional embedding in 3D space).
Such solutions play an important role in the realistic modelling of cosmic structure formation, since, in the fundamental coordinate system of a given collapsing patch, shell-crossings generically arise  as an almost one-dimensional phenomena with the formation of pancake-like structures~(e.g.~\cite{Melott:1992vp}).

\paragraph{Secondary gravitational infall.} 
Once matter streams have crossed for the first time, nearby particles experience non-trivial accelerations due to the secondary gravitational infall into the potential well.  Subsequently, a self-trapped quasi-stationary structure emerges (e.g., a dark-matter halo in the long-term), forcing distinct matter streams to cross again -- the second shell-crossing. For topological reasons, the number of streams generically bifurcates from three to five at the second shell-crossing (and then from five to seven at the third shell-crossing; etc.).

\begin{figure}
  \begin{center}
  
  \includegraphics[width=\textwidth]{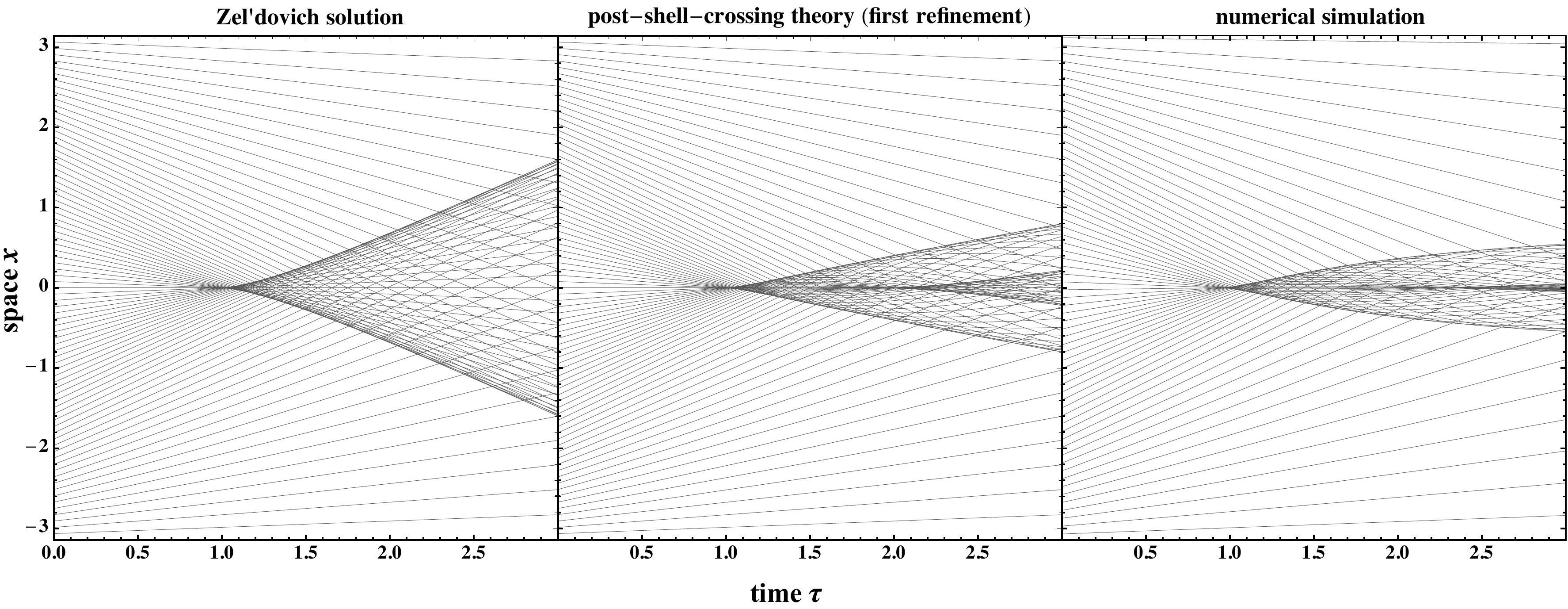}

  \end{center}
\caption{Evolution of selected dark-matter trajectories, using the same 1D setup as in Fig.\,\ref{fig:map-density}. {\bf Left panel:}  The analytical Zel'dovich solution~\cite{Zeldovich:1969sb} predicts the first shell-crossing (here at $\tau=1$ and $x=0$), however not the second one. Instead, non-linear structures are incorrectly washed out after the first shell-crossing. {\bf Central panel:}  By contrast, modern post-shell-crossing theories (here the purely analytical result of \cite{Rampf:2019nvl})  predict the second shell-crossing (here at $\tau \simeq 2.2$ and $x=0$), however would require higher refinements to predict the third shell-crossing (not shown). {\bf Third panel:} Fully numerical result (courtesy O.\,Hahn) employing the same algorithm as for Fig.\,\ref{fig:phasespace-color}, where the second shell-crossing occurs at $\tau \simeq 2.4$. 
}
\label{fig:trajectories}
\end{figure}

While numerical simulations revealed the second shell-crossing several decades ago (e.g.~\cite{1980MNRAS.192..321D,Fillmore:1984wk}), it was not predicted by theoretical means until recently,
when S.~Colombi was able to accurately estimate the involved gravitational forces in multi-streaming regions from first principles~\cite{Colombi:2014lda} (see also \cite{Rampf:2019nvl,Taruya:2017ohk,Pietroni:2018ebj}).
This is shown in Fig.\,\ref{fig:trajectories}, where we compare the classical Zel'dovich solution (left panel) against the refined theoretical prediction of particle trajectories (central panel) and a numerical simulation (right panel). 
First notice that in regions with no multi-streaming, the trajectories in all panels are identical and simply straight lines, indicating that particles follow their initially prescribed constant velocity with zero acceleration (note this feature of exact ballistic motion in single-stream regions is only true in the one-dimensional case). 

Starting at shell-crossing (here at $\tau=1$), some particle trajectories in multi-streaming regions are bent back into the center of the potential well, a feature that is not encapsulated in Zel'dovich's theory, but is captured in refined theoretical predictions as well as in the numerical simulation (central and right panel of Fig.\,\ref{fig:trajectories}).

Predicting the third and even higher shell-crossings by analytical means is in principle feasible with straightforward extensions of the analytical techniques of~\cite{Rampf:2019nvl,Colombi:2014lda,Taruya:2017ohk}, but  have not been performed so far; this is also the reason why the shown refined theoretical prediction in Fig.\,\ref{fig:trajectories} does not agree with the numerical prediction at later times.

Of course, the actual Universe with random initial conditions is more complicated than the depicted scenario in Fig.\,\ref{fig:trajectories}. In particular, 
due to the random nature, collapsing structures  are generally not isolated objects. 
Instead, structures are formed on smaller spatial scales that eventually merge to combined structures  on larger scales  (e.g., dark-matter halos). Resolving such merging events is currently beyond the reach of analytical considerations, and only amenable by employing numerical simulations (but see \cite{Taruya:2017ohk,Pietroni:2018ebj} for related semi-analytical avenues using adaptive smoothing). On top of that, the process of halo or galaxy formation requires  accurate physical modelling that goes well beyond the cosmic Vlasov--Poisson equations, in particular physics from visible (baryonic) matter; see e.g.\ \cite{Schaye:2014tpa,Vogelsberger:2014dza,Springel:2017tpz,Arico:2019ykw,Lewandowski:2014rca}.

In the following sections, we will focus on the mathematical analysis of the dark-matter evolution before and (shortly) after the first shell-crossing -- a regime that until very recently was not accessible by analytical means. Nonetheless, in sections~\ref{sec:effective}--\ref{sec:num} we will summarise complementary theoretical (effective) approaches as well as numerical simulations  that consider the dark-matter evolution at substantially later times.

%%%%%%%%%%%%%%%%%%%%%%%%%%%%%%
\section{Master equations and solution strategy}\label{sec:mastereqs}

\subsection{Lagrangian evolution equations}

The Vlasov--Poisson equations~\eqref{eq:VPLag} are not yet in a form that allows for straightforward analytical considerations. One reason for this is the appearance of the Eulerian space derivative with respect to the coordinate $\fett{x}$, which in Lagrangian coordinates is not an independent variable. 
Indeed, earlier we have introduced the Lagrangian map $\fett{q} \mapsto \fett{x}(\fett{q},\tau)$ which implies for the conversion of derivatives  that $\nabla_{q_i} = (\partial x_j/\partial q_i) \nabla_{x_j}$, where, from here on, Latin indices $i,j, \ldots =1,2,3$ denote the three spatial components, and (Einstein) summation over repeated indices is assumed.

To proceed, one introduces the {\it Lagrangian displacement field}  $\fett{\xi}(\fett{q},\tau)$, 
\be
  \fett{\xi}(\fett{q},\tau) := \fett{x}(\fett{q},\tau) -\fett{q} ,
\ee
which  encodes the whole dynamical information of the Vlasov--Poisson equations for cold dark matter -- at all times. Indeed, once $\fett{\xi}$ is determined, one can retrieve the matter density (using Eq.\,\ref{eq:mass}), the velocity (possibly involving multi-stream averaging),   and higher kinetic moments of the dark-matter distribution function, such as the velocity dispersion tensor (see e.g.\ \cite{Buehlmann:2018qmm}).

Introducing the operator $\hat {\cal T}_n := (2\tau^2/3) \partial_\tau^2 + \tau \partial_\tau - n$
and applying suitable (derivative) operations to the Vlasov--Poisson equations~\eqref{eq:VPLag}, summarised below, the Lagrangian evolution equations for the divergence and curl of the displacement are
\begin{empheq}[box=\widefbox]{align} \label{eqs:master}
 \hat {\cal T}_{1} \, \Big[\nabq \cdot  \fett{\xi} \Big] &= {\cal W} + {\cal M} \, ,
\qquad \hat {\cal T}_0 \Big[ \nabq \times \fett{\xi} \Big] =  \left( \hat {\cal T}_0 \nabq   \xi_l \right) \times \nabq  \xi_l \,,
\end{empheq}
which, once solved respectively for $\nabq \cdot  \fett{\xi}$ and $\nabq \times \fett{\xi}$, yield the formal solution for the displacement using a Helmholtz decomposition ($\nabla_{\fett{q}}^{-2}$ is the inverse Laplacian),
\be \label{eq:Helm}
  \fett{\xi} =  \nabla_{\fett{q}}^{-2} \Big(  \nabq \left[ \nabq \cdot \fett{\xi} \right] - \nabq \times \left[ \nabq \times \fett{\xi} \right]  \Big) .
\ee
We have defined two non-linear source terms (see below for physical interpretations)
\begin{align}
   {\cal W} &=   \left[  \xi_{j,i} - \xi_{l,l} \delta_{ij} \right] \hat {\cal T}_{\nicefrac 1 2\,} \xi_{i,j} - \tfrac 1 2 \varepsilon_{ikl} \varepsilon_{jmn} \xi_{k,m} \xi_{l,n} \hat  {\cal T}_{\nicefrac 1 3\,} \xi_{i,j} \,, \label{eq:calW} \\
  {\cal M} &= 1    - \det[\nabq \fett{x}]  \int \delta_{\rm D} \left[ \fett{x}(\fett{q},\tau)- \fett{x}(\fett{q}',\tau) \right] \, \dd^3 q' \,, \label{eq:calM}
\end{align}
where ``$_{,i}$'' denotes a partial derivative with respect to Lagrangian component $q_i$, while $\varepsilon_{ijk}$ is the fundamental anti-symmetric tensor and $\delta_{ij}$ the Kronecker delta. 
Here and in the following, displacements $\fett{\xi}$ and current positions $\fett{x}$ depend on $\fett{q}$ and $\tau$ if not otherwise stated.
Equations~\eqref{eqs:master}--\eqref{eq:calM} are derived for an Einstein--de Sitter cosmological model,
but are equally valid for a $\Lambda$CDM Universe upon identifying $\tau$ with the standard 
$\Lambda$CDM linear growth function $D$ and the replacement 
$\hat {\cal T}_n \to \hat {\cal T}_n^\Lambda = (2D^2/3) \partial_D^2 + g D \partial_D - n g$, where  
$g= (D/\partial_t D)^2 a^{-3}$.

\paragraph{Derivation of equations~\eqref{eqs:master}.}
To derive the divergence equation, one first takes the Eulerian space derivative of the Vlasov equation~\eqref{eq:VPLag}, thereby allowing one to express its right-hand side by the Poisson equation and the mass conservation~\eqref{eq:mass}. Next, one converts the remaining Eulerian space derivative according to $\nabla_{x_i} = (\partial q_j/\partial x_i) \nabla_{q_j}$, where $\partial q_j/\partial x_i =  \varepsilon_{ikl} \varepsilon_{jmn}  x_{k,m}  x_{l,n}/(2 \det[\nabq \boldsymbol x])$. Multiplying this resulting equation by $(\tau^2\det[\nabq \fett{x}]/3)$ and using the exact identity
 $\det[\nabq \fett{x}] = 1 + \xi_{l,l} + (1/2)  [\xi_{l,l} \xi_{m,m} - \xi_{l,m} \xi_{m,l} ] + (1/6) \varepsilon_{ikl} \varepsilon_{jmn} \xi_{k,m} \xi_{l,n} \xi_{i,j}$, one then obtains the desired result (see e.g.~\cite{Matsubara:2015ipa,Buchert:1987xy} for equivalent derivations for the case ${\cal M}=0$).
To derive the equation on the right in~\eqref{eqs:master}, one first multiplies~\eqref{eq:VPLag} by $2\tau^2/3$, leading to $\hat {\cal T}_0 x_l = - \tau \nabla_{x_l} \varphi$. Multiplying the last equation 
by the matrix element $\partial x_l/ \partial q_j$ then yields
$ x_{l,j}  \hat {\cal T}_0 x_l = - \tau \nabla_{q_j} \varphi$. Finally, taking the Lagrangian curl of this leads to 
$(\nabq x_l) \times  \hat {\cal T}_0 \nabq x_l = \fett{0}$ \cite{Matsubara:2015ipa,Zheligovsky:2013eca},
which, once written for the displacement $\fett{\xi}= \fett{x}-\fett{q}$, leads directly to the second equation in~\eqref{eqs:master}.

\paragraph{Physical interpretation of source terms.}
${\cal M}$ is only nonzero once there are locations with an overlap of multiple matter streams or,  stated the other way around, ${\cal M}$ is exactly zero everywhere, as long as the flow is still single stream (mono-kinetic). Indeed, for single-stream flows, the Lagrangian mapping $\fett{q} \mapsto \fett{x}$ is injective with $\det [\nabq \fett{x}] >0$ (corresponding to $\nabla_q x>0$ in 1D; cf.~Fig.\,\ref{fig:map-density}), implying that $\fett{x}(\fett{q})-\fett{x}(\fett{q}')$ has only a single root at $\fett{q}'=\fett{q}$. Consequently, $\int \delta_{\rm D}\left[ \fett{x}(\fett{q},\tau)- \fett{x}(\fett{q}',\tau) \right] \, \dd^3 q' = 1/\det [\nabq \fett{x}(\fett{q})]$ and thus ${\cal M} = 0$.

By contrast, ${\cal W}$ is the result of converting Eulerian space derivatives to Lagrangian space, thereby accumulating quadratic and cubic terms in the displacement. With regard to the physical interpretation: 
for generic cosmological initial conditions in three space dimensions,
${\cal W}$ constitutes a sub-dominant (perturbatively suppressed) term  during the very early gravitational evolution. Thus, the first evolution equation in~\eqref{eqs:master} becomes $\hat {\cal T}_{1} \left[ \nabq \cdot  \fett{\xi} \right] \simeq 0$ at sufficiently early times in the single-stream regime. This is an ordinary differential equation in time for $\nabq \cdot \fett{\xi}$ and, importantly, is a linear equation in the displacement. Supplemented with suitable initial conditions, the solution of this differential equation yields the aforementioned Zel'dovich approximation (see the following section).
Furthermore, ${\cal W}$ is exactly zero at all times, iff the initial data depends only on a single-space coordinate (since derivatives in the transverse flow direction vanish in that case): This is the technical reason why the Zel'dovich theory, a linear description in Lagrangian coordinates, is exact in 1D until shell-crossing.

Finally, if the vorticity $\fett{w} := \nabx \times \fett{v}$  vanishes initially,  as assumed implicitly when using the boundary conditions~\eqref{eq:slaving}, then the second equation in~\eqref{eqs:master} amounts to a conservation equation for the zero-vorticity condition. This equation and its variants have a long history tracing all the way back to Cauchy in 1815; see \cite{Zheligovsky:2013eca} for details and e.g.~\cite{Matsubara:2015ipa,Ehlers:1996wg,Rampf:2012up} for contemporary formulations in cosmology.

\subsection{Theoretical solution strategy} \label{sec:strategy}

Obtaining theoretical solutions for the Vlasov--Poisson equations essentially require three ingredients as summarised below.
\begin{enumerate}[leftmargin=20pt, itemsep=5pt]
 \item[\circled{1}] {\bf Linear analysis ($\boldsymbol{{\cal W} = 0}$, $\boldsymbol{{\cal M}=0}$).} Linearising the evolution equations~\eqref{eqs:master} around the steady state of the displacement $\fett{\xi}(\fett{q},\tau)=\fett{x}(\fett{q},\tau)-\fett{q}$, one obtains
\begin{align} 
   \hat {\cal T}_{1} \, \Big[\nab \cdot  \fett{\xi} \Big] =0 \,,  \qquad \qquad \qquad \hat {\cal T}_0 \Big[ \nab \times \fett{\xi} \Big] = \fett{0} \,, 
\end{align}
where from here on $\nab := \nabq$, and we remind the reader that $\hat {\cal T}_n =(2\tau^2/3) \partial_\tau^2 + \tau \partial_\tau - n$.  The general solutions of these equations are respectively
\be \label{eq:linearSols}
  \nab \cdot  \fett{\xi} = \lambda \, \tau + \mu \, \tau^{-3/2} \,, \qquad \qquad   \nab \times \fett{\xi}  = \fett{\omega} +  \fett{\chi} \tau^{-1/2} \,,
\ee
where $\lambda,\, \mu,\, \fett{\omega}$ and $\fett{\chi}$ are spatial constants to be determined by the boundary conditions (the first of the solutions in~\ref{eq:linearSols} is formally identical with the one for the linearised density contrast; see e.g.~\cite{Bernardeau:2001qr,Peebles:1980}).
If the boundary conditions~\eqref{eq:slaving} are employed, then only the growing solution $\sim \tau$ is selected and one arrives at the classical Zel'dovich solution \cite{Zeldovich:1969sb,Buchert:1989xx}
\be \label{eq:ZA}
  \fett{x}(\fett{q},\tau) = \fett{q} + \tau \,\fett{v}^{\rm ini} \,,   \qquad \qquad  \fett{v}^{\rm ini}(\fett{q}) = - \nab \varphi^{\rm ini}(\fett{q}) \,, 
\ee
where $\varphi^{\rm ini}$ is the initial gravitational potential.
Of course, other solutions (e.g., with terms that decay in~$\tau$) exist as well \cite{Buchert:1992ya,Bouchet:1994xp,Nadkarni-Ghosh:2012byn}, but require a more sophisticated analysis of the boundary layer (see e.g.~\cite{Fidler:2016tir,Adamek:2017grt}).

 \item[\circled{2}] {\bf Shell-crossing solutions ($\boldsymbol{{\cal W} \neq 0$, ${\cal M}=0}$).} As long as the flow is still single-stream, the evolution equations for the particle displacements are exactly
\be \label{eqs:masterSingle}
  \hat {\cal T}_{1} \, \Big[\nab \cdot  \fett{\xi} \Big] = {\cal W} \, ,
\qquad \qquad  \hat {\cal T}_0 \Big[ \nab \times \fett{\xi} \Big] =   \left( \hat {\cal T}_0 \nab  \xi_l \right) \times \nab  \xi_l \,.
\ee
If we impose the boundary conditions~\eqref{eq:slaving}, then these equations remain analytic for $\tau \to 0$ \cite{Brenier:2003xs,Zheligovsky:2013eca},  thereby suggesting that a series representation for the displacement in powers of $\tau$ (the linear growth time), around $\tau=0$, is amenable:
\begin{subequations} \label{eqs:LPT}
\be \label{eq:Ansatz}
   \fett{\xi}(\fett{q},\tau) = \sum_{n=1}^\infty \fett{\zeta}^{(n)}(\fett{q}) \,\tau^n \,.
\ee
Here, $\fett{\zeta}^{(1)}(\fett{q}) = -\nab \varphi^{\rm ini}$ is the first of infinitely many,
purely space-dependent Taylor coefficients. We note that this factorisation into spatial and temporal parts 
persists also for a $\Lambda$CDM Universe \cite{Ehlers:1996wg}.
Plugging~\eqref{eq:Ansatz} into~\eqref{eqs:masterSingle}, one obtains explicit all-order 
recursion relations \cite{Rampf:2015mza,Matsubara:2015ipa,Zheligovsky:2013eca,Rampf:2012up,Rampf:2020hqh,Schmidt:2020ovm}
\begin{align}
  \nab \cdot \fett{\zeta}^{(n)} &=  -\nab^2 \varphi^{\rm ini} \delta^n_1 + \sum_{0<s<n} \tfrac{(3-n)/2-s^2-(n-s)^2}{(2n+3)(n-1)} \left[   \zeta_{i,i}^{(s)} \zeta_{j,j}^{(n-s)} - \zeta_{i,j}^{(s)} \zeta_{j,i}^{(n-s)} \right] \nonumber \\
  &\quad + \sum_{a+b+c=n} \tfrac{(3-n)/2 -a^2-b^2 -c^2}{3(2n+3)(n-1)}\varepsilon_{ikl} \varepsilon_{jmn} \zeta_{k,m}^{(a)} \zeta_{l,n}^{(b)}  \zeta_{i,j}^{(c)} =: L^{(n)} \,, \label{eq:scalarRec} \\
  \nab \times \fett{\zeta}^{(n)} &= \tfrac 1 2 \! \sum_{0<s<n} \tfrac{n-2s}{n}\, \nab \zeta_l^{(s)} \times \nab \zeta_l^{(n-s)} =: \fett{T}^{(n)}\,, \label{eq:vectorRec}
\end{align}
\end{subequations}
from which $\fett{\zeta}^{(n)}= \nabla^{-2}(\nab L^{(n)}- \nab \times \fett{T}^{(n)})$ follows,  
and subsequently~$\fett{\xi}=\sum_n\fett{\zeta}^{(n)}\,\tau^n$.
Note that  for sufficiently smooth $\varphi^{\rm ini}$,
as it is the case for cosmological initial conditions, 
there exists mathematical proofs of time-analytic displacements until some finite time \cite{Rampf:2015mza,Zheligovsky:2013eca}, as well as numerical evidence of convergence at shell-crossing and even shortly after~\cite{Rampf:2020hqh}; see section~\ref{sec:sc-random} for details.

Once the displacement is determined to sufficient accuracy, the time of first shell-crossing, denoted with $\tau_{\star}$, can be determined by searching for the particle $\fett{q}=\fett{q}_{\star}$ for which the density blows up for the first time (cf.\ equation~\ref{eq:mass2}), i.e., 
\be \label{eq:SCcondition}
 \delta(\fett{x}(\fett{q}_{\star}, \tau_{\star})) = \frac{1}{\det[\nab \fett{x}(\fett{q},\tau_{\star})]}\bigg|_{\fett{q}=\fett{q}_{\star}} -1 = \infty \,.
\ee 
Of course, not only the particle with label $\fett{q}_{\star}$ but all particles can be evolved until~$\tau_{\star}$; in the following we denote the family of displacements at $\tau_{\star}$ with $\fett{\xi}_{\!\!{\star}} := \fett{\xi}_{\!\!{\star}}(\fett{q},\tau=\tau_{\star}) =\sum_{n=1}^{n_{\rm max}} \fett{\zeta}^{(n)}(\fett{q})\, \tau_{\star}^n$, truncated at sufficiently large order~$n_{\rm max}$.

 \item[\circled{3}] {\bf Post-shell-crossing solutions ($\boldsymbol{{\cal W} \neq 0$, ${\cal M}\neq0}$).}
Instantly after the fluid bifurcates at shell-crossing time $\tau=\tau_{\star}$ to three streams at location $\fett{q}= \fett{q}_{\star}$, the multi-stream source term ${\cal M}$ is nonzero in the neighbourhood of $\fett{q}_{\star}$ (blue shaded region in  Fig.\,\ref{fig:map-density}). 
In this case, the evolution equation for $\nab \cdot  \fett{\xi}$ at $\tau \geq \tau_{\star}$ can be formally solved by using the method of variation of constants, i.e.,
\begin{subequations} \label{eqs:formalPSCsolution}
\be
  \hat {\cal T}_{1} \, \Big[\nab \cdot  \fett{\xi} \Big] = {\cal W} + {\cal M} \,, \qquad \Rightarrow \quad  \nab \cdot \fett{\xi} = \lambda(\tau)\, \tau + \mu(\tau)\, \tau^{-3/2} \,,
\ee
with
\begin{align} 
 \lambda(\tau) &=  \tfrac{3}{5} \tau_{\star}^{-1} \nab \cdot  \xistar +\tfrac 2 5 \nab \cdot \dotxistar + \tfrac{3}{5} \int_{\tau_{\star}}^\tau  \eta^{-2} \left[ {\cal W}(\eta)+{\cal M}(\eta) \right]\, \dd \eta \,, \\
 \mu(\tau) &= \tfrac 2 5 \tau_{\star}^{3/2} \nab \cdot \left(  \xistar - \tau_{\star}  \dotxistar  \right)
   - \tfrac{3}{5}  \int_{\tau_{\star}}^\tau  \eta^{1/2} \left[{\cal W}(\eta)+{\cal M}(\eta) \right]\, \dd \eta \,,
\end{align}
\end{subequations}
where $\dotxistar = \partial_\tau \fett{\xi}_{\!\!\star}(\fett{q},\tau)|_{\tau =\tau_\star}$; 
see \cite{Rampf:2019nvl,Colombi:2014lda,Taruya:2017ohk,Pietroni:2018ebj} for complementary methods applied to the one-dimensional case.
Note that  the re-occurring terms ${\cal W}$ and ${\cal M}$ in~\eqref{eqs:formalPSCsolution} are \mbox{\it a priori} unknowns for times $\tau > \tau_{\star}$, since they are functions of the unknown post-shell-crossing displacement~$\fett{\xi}$ (cf.\ equations~\ref{eq:calW}--\ref{eq:calM}).
However, the flow directions of the matter stream are known until shell-crossing exactly --- and at least approximatively shortly after, essentially by an argument of momentum conservation. Thus,  
the unknowns ${\cal W}$ and ${\cal M}$ in~\eqref{eqs:formalPSCsolution} can be approximatively determined 
by extrapolating the shell-crossing solutions to times shortly after shell-crossing. 
Specifically, for $\tau > \tau_{\star}$, we approximate
\be \label{eqs:LOapprox}
  {\cal W}(\fett{x}(\fett{q},\tau)) \simeq {\cal W}(\fett{x}_{\!\fett{\bullet}}(\fett{q},\tau)) \,,   
 \qquad   
 {\cal M}(\fett{x}(\fett{q},\tau)) \simeq {\cal M}(\fett{x}_{\!\fett{\bullet}}(\fett{q},\tau)) \,, 
\ee
in the first iteration, where 
$\fett{x}_{\!\fett{\bullet}} (\fett{q},\tau) := \fett{q}+ \sum_{n=1}^{n_{\rm max}} \fett{\zeta}^{(n)}(\fett{q})\,\tau^n$ is the  truncated shell-crossing displacement until $n=n_{\rm max}$ with Taylor coefficients $\fett{\zeta}^{(n)}$, the latter determined through~\eqref{eq:scalarRec}--\eqref{eq:vectorRec}. 
We remark that the computation of the multi-streaming source ${\cal M}(\fett{x}_{\!\fett{\bullet}})$ is still non-trivial and requires normal-form considerations inherited 
from catastrophe theory \cite{Berry_1977,Arnold1980}; see section~\ref{sec:PSC} for details.

As first observed by S.\,Colombi \cite{Colombi:2014lda}, 
the resulting dark-matter displacements subject to the leading-order approximation for ${\cal M}$ are fairly accurate,
and already overcome the major obstacle of theoretically predicting the second shell-crossing.
Suitable higher-order refinements for resolving the regime between the first and second shell-crossing are straightforward~\cite{Rampf:2019nvl,Taruya:2017ohk,Pietroni:2018ebj}; so are applications beyond~1D (e.g., \cite{Chen:2020zuf,Zimmermann:2021sqo}).
Finally, predicting the third shell-crossing by theoretical means is in principle possible, but increasingly challenging. 
For this one would first need to provide boundary conditions at the second shell-crossing, followed by approximating ${\cal W}$ and~${\cal M}$ along the lines as summarised above. 

\end{enumerate}

%%%%%%%%%%%%%%%%%%%%%%%%%%%%%%%%%%%
\section{Results before and until shell-crossing (\texorpdfstring{${\cal M}=0$}{M=0})}

We first provide an overview of approximative solution techniques, valid as long as multi-streaming has not yet occurred. Then, in section~\ref{sec:sc-anal} and~\ref{sec:sc-random},  we investigate the mathematics of shell-crossings respectively for simplified and random initial conditions.

\subsection{Perturbative methods and overview of applications} \label{sec:pertZA++}

Approximative techniques to Vlasov--Poisson have a long history in cosmology. Most of them rely on the single-stream description of the Vlasov--Poisson equations, which, by applying kinetic moments to~\eqref{eq:VP}, reduce to a system of fluid equations with vanishing velocity dispersion (see e.g.~\cite{Bernardeau:2001qr} for details). 
The basic motivation for using such perturbative techniques is, that the larger and/or the earlier the cosmological scales considered, the more these techniques deliver physically meaningful results. 
This is so for at least two reasons, namely  that (1) on sufficiently large scales and/or early times, the matter evolution follows mostly the overall Hubble flow due to the expanding Universe, which justifies generic expansions around linearised field variables, and that (2) on such spatio-temporal scales, shell-crossing effects  are yet negligible.
The overall literature is vast, therefore we attempt here to only summarise some of the rather recent applications.

\paragraph{Eulerian methods.} Instead of seeking a power series representation for the displacement in Lagrangian coordinates, one can as well perturbatively expand the fluid variables in Eulerian coordinates~$\fett{x}$; for example the density may be represented as a power series in the linear growth time, i.e., $\delta(\fett{x},\tau)= \sum_{n=1}^\infty \delta^{(n)}(\fett{x})\, \tau^n$, where the first Taylor coefficient is $\delta^{(1)}(\fett{x})=\nabla^2 \varphi^{\rm ini}$. The general framework for this is called Eulerian or standard perturbation theory (SPT) \cite{Peebles:1980,Fry:1983cj,Goroff:1986ep,Bertschinger:1993zv,Blas:2013aba}, and there exists  explicit all-order recursion relations~\cite{Goroff:1986ep} as well as a variety of SPT extensions (e.g.~\cite{Crocce:2005xy,Pietroni:2008jx,Bernardeau:2008fa,Anselmi:2010fs,Bernardeau:2011vy,Crocce:2012fa,Carlson:2012bu,Kitaura:2012tj,Vlah:2014nta}); see also \cite{Bernardeau:2001qr} for an extensive review. A straightforward application of SPT is to determine so-called loop corrections to $n$-point statistics of the matter density contrast, e.g., to the matter power- and bispectrum which are the Fourier transforms of the 2-point and 3-point correlation functions of $\delta$, respectively \cite{Suto:1990wf,Jain:1993jh,Scoccimarro:1997st,Nishimichi:2008ry,Carlson:2009it,Taruya:2012ut,Simonovic:2017mhp}. Furthermore, certain effective approaches based on SPT are capable of incorporating shell-crossing effects to some extent; see section~\ref{sec:effective} for details.
Finally, in Eulerian perturbative approaches, physics beyond Vlasov--Poisson can be straightforwardly implemented. One important example for this relates to the biasing problem, where one seeks functional relationships between the spatial distributions of the dark-matter and visible (baryonic) density; see e.g.\ \cite{Matsubara:2011ck,Desjacques:2016bnm,Schmittfull:2018yuk,Eggemeier:2021cam}.

We remark that, so far, the mathematical convergence of the SPT series %until shell-crossing 
has not been explicitly demonstrated, except for the simplified cases of one-dimensional~\cite{McQuinn:2015tva} and spherical collapse \cite{Rampf:2017tne}.
Nonetheless, convergence for cosmological initial conditions until -- but excluding --  shell-crossing is likely, mainly as a consequence of the analyticity results in Lagrangian coordinates \cite{Rampf:2020hqh,Schmidt:2020ovm}. However, the  speed of convergence of the power series for the density is expected to be substantially slower as compared to the series for the displacement in the Lagrangian case, due to the presence of the convergence limiting density singularity at shell-crossing.

\paragraph{Zel'dovich approximation.} 
One of the most important yet simple theoretical models of cosmic structure formation is the Zel'dovich approximation, which is the linear solution of the Vlasov--Poisson equations in Lagrangian coordinates with 3D initial conditions (taking the boundary conditions~\ref{eq:slaving} into account).
Specifically, the Zel'dovich approximation is achieved by keeping only the $n=1$ contribution in the infinite Taylor series~\eqref{eq:Ansatz} for the displacement. 
As we have elucidated above, the Zel'dovich approximation essentially boils down to a  ballistic description for dark-matter trajectories, which we repeat here for convenience,
\be \label{eq:ZArep}
  \fett{x}(\fett{q},\tau) = \fett{q} + \tau \,\fett{v}^{\rm ini}(\fett{q}) \,,   \qquad \text{with} \qquad  \fett{v}^{\rm ini}(\fett{q}) = - \nab \varphi^{\rm ini}(\fett{q}) \,.
\ee
Although being a linear approximation to Vlasov--Poisson in Lagrangian coordinates, it leads to a highly non-linear prediction for the density in Eulerian coordinates, essentially due to the inherent non-linearity in the mapping $\fett{q} \mapsto \fett{x}$,
thereby providing a surprisingly accurate description of the non-linear gravitational collapse.
Indeed, since for~\eqref{eq:ZArep} we have $\det[\nabq \fett{x}] = \det[\delta_{ij} - \tau \varphi_{,ij}^{\rm ini}]$, the mass conservation law~\eqref{eq:mass2} can be written as \cite{Zeldovich:1969sb}
\be \label{eq:densityZA}
  \delta(\fett{x}(\fett{q},\tau)) +1  = \big[(1 - \tau \lambda_1(\fett{q})) (1 - \tau \lambda_2(\fett{q})) (1 - \tau \lambda_3(\fett{q})) \big]^{-1}
\ee
in the fundamental coordinate system spanned by $\nabq \fett{x}$, where $\lambda_{1,2,3}$ are the eigenvalues of the Hessian of the initial gravitational potential. From the relation~\eqref{eq:densityZA} it is clear that shell-crossing is reached when any one of the three eigenvalues goes to zero. We remark that the instantaneous vanishing of two or even three eigenvalues is essentially excluded for random initial conditions \cite{Doroshkevich1970}.

The Zel'dovich approximation has several key advantages over the corresponding linear solution in Eulerian coordinates, where the latter predicts that $\delta(\fett{x},\tau)=  \tau \nabla^2 \varphi^{\rm ini}$ at first order in SPT (see above). Evidently, first-order SPT provides a completely unrealistic prediction for the time of first shell-crossing which is achieved only at $\tau \to \infty$ for sufficiently smooth initial conditions. By contrast, we know by now that the first shell-crossing occurs at times $\tau = \tau_\star \ll 1$  for cosmological initial conditions (the precise time depends on spatial resolution; see section~\ref{sec:sc-random}), while the Zel'dovich approximation predicts $\tau_\star$ to about 20\% accuracy. 
Another advantage of the Zel'dovich approximation is that it correctly predicts the emergence of a three-stream regime in the phase-space after shell-crossing (see e.g.~Fig.~\ref{fig:trajectories}), which in SPT is never achieved, even at arbitrary high orders.

Important (and rather recent) applications of the Zel’dovich approximation and higher-order refinements include:
\begin{itemize}
  \item Generating initial conditions for cosmological simulations \cite{KlypinShandarin1983,Efstathiou:1985re,Scoccimarro:1997gr,Valageas:2001qe};  see also \cite{Michaux:2020yis,Buchert:1993df,Crocce:2006ve} for higher-order refinements within the framework of Lagrangian perturbation theory \cite{Buchert:1992ya,Ehlers:1996wg,Buchert:1989xx,Bouchet:1994xp,Moutarde1991,Bouchet:1992xz,Buchert:1993xz,Buchert:1993ud,Rampf:2012xa} (LPT), which are predictions based on perturbative truncations of equation~\eqref{eqs:LPT}. 
  \item Predicting the change of photon momenta emitted from observed luminous tracers of dark matter, due to peculiar motion and the expansion of space (``redshift-space distortions'')  \cite{Matsubara:2007wj,Taylor:1996ne,Chen:2020zjt,Valogiannis:2019nfz,Jackson:1971sky,Kaiser:1987qv,Fisher:1995ec}.

 \item Theoretically modelling the spatial distribution of cosmic neutral hydrogen through its 21cm emission line \cite{McDonald:1998pm,Hui:1998vf,Jasche:2012kq,Villaescusa-Navarro:2018vsg,Modi:2019hnu}.

 \item Investigating topological features of the cosmic large-scale structure, such as clusters, sheets, and filaments (cf. Fig.\,\ref{fig:LSS}) \cite{Bond:1995yt,Hahn:2006mk,Aragon-Calvo:2010xbo,Falck:2012ai,Hoffman:2012ft,Leclercq:2013kza,Feldbrugge:2017ivf}.

 \item Modelling the baryonic acoustic oscillation feature imprinted in statistics of cosmic structures \cite{Eisenstein:2006nk,Crocce:2007dt,White:2015eaa,Chen:2019lpf,Chen:2020fxs,Aviles:2020wme,Tamone:2020qrl}; and

 \item improving statistical predictions of the large-scale structure by pairing/emulating techniques with numerical simulations \cite{Tassev:2013pn,Howlett:2015hfa,Feng:2016yqz,Chartier:2020pmu,Kokron:2021xgh,Zennaro:2021bwy,Arico:2021izc}.

\end{itemize}

\subsection{Analytical shell-crossing solutions for simplified initial conditions}\label{sec:sc-anal}

Even without multi-streaming (${\cal M}=0$ in equation~\ref{eqs:master}), the Vlasov--Poisson equations are highly non-linear, especially near shell-crossings where \mbox{$\delta \to \infty$.}
Nonetheless, there are a few purely analytical shell-crossing solutions.
Analytical solutions play an important role in theoretical cosmology, and can be used as  test problems for state-of-the-art simulation techniques (e.g., \cite{Vogelsberger:2008qb,Abel:2011ui,Shandarin:2011jv,Hahn:2012ma,Hahn:2015sia,Sousbie:2015uja,Stucker:2019txm}).

\paragraph{Quasi-one dimensional collapse.} 
The Zel'dovich solution~\eqref{eq:ZArep} becomes exact until shell-crossing, if the initial data depends only on one space coordinate, since in that case the displacement can only depend on the same space coordinate, but not on the others; hence
${\cal W} =0$ (see discussion after equation~\ref{eq:calW}). Figuratively, the one-dimensional problem embedded in 3D can be thought of as stacked mass sheets/curtains along the coordinate direction.

Suppose now that the initial data depends mostly on one coordinate, say $q_1$, but also depends weakly on the other coordinates. Mathematically and actually without loss of generality, the corresponding initial gravitational potential can be set to
\begin{subequations}
\be
 \label{eq:Q1Dics}
  \varphi^{\rm ini} = -\cos q_1 + \epsilon \phi^{\rm ini}(q_1,q_2,q_3) \,,
\ee
where $\epsilon >0$ is a perturbation parameter that controls the smallness of the deviation from the one-dimensional problem, while $\phi^{\rm ini}$ is an arbitrary $2\pi$-periodic function. This is thus an intrinsically three-dimensional problem; within the above picture, the mass sheets have now a weak functional dependence in all coordinate directions.

\begin{figure}
  \begin{center}
  
  \includegraphics[width=\textwidth]{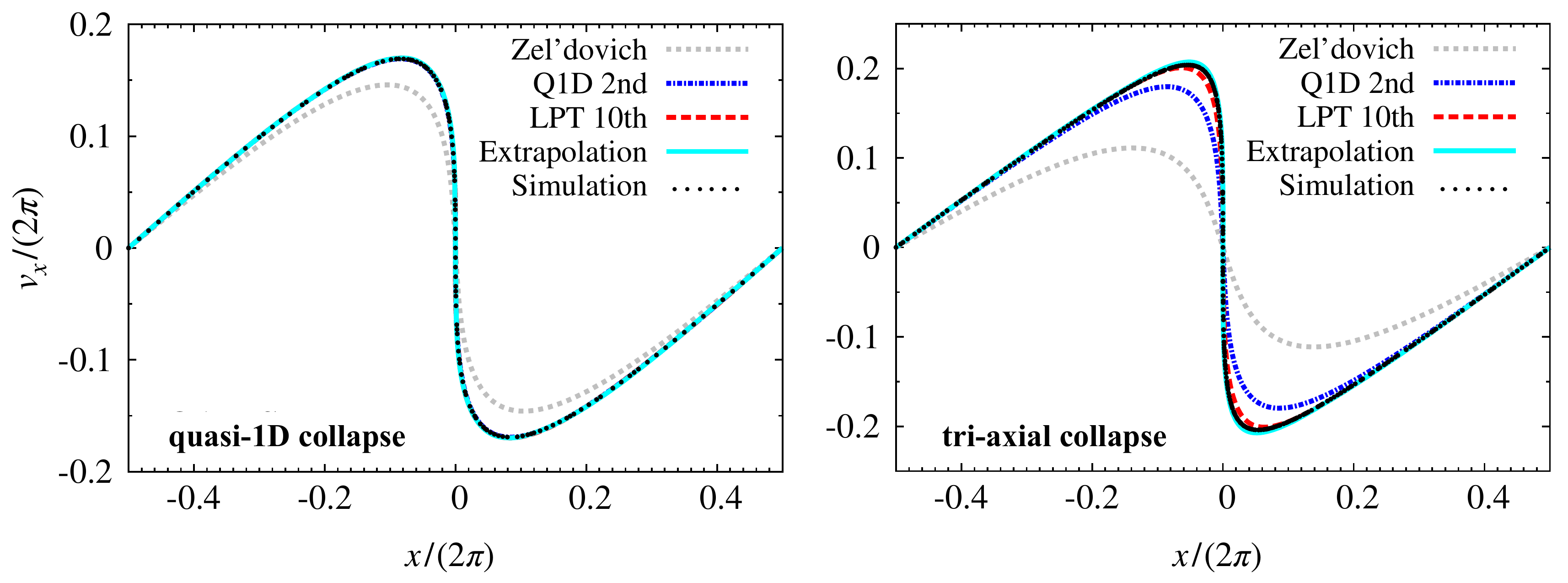}

  \end{center}
\caption{Comparison of various analytical shell-crossing predictions against numerical simulations (black dotted lines) on the torus with period $2\pi$. The blue (dashed dotted) line denote results from the quasi-one-dimensional theory at second order (equation~\ref{eq:quasi1D}), while the red (dashed) lines are obtained by truncating the Taylor series~\eqref{eqs:LPT} up to the 10th order (denoted with LPT). The cyan (solid) lines are fits obtained from an asymptotic extrapolation technique of the 10th-order LPT results.
The left panel shows the quasi-1D collapse with $x$ being the dominant flow direction, while the initial gravitational potential for the right panel are three crossed sine-waves with different amplitudes. Evidently, LPT exemplifies convergent behaviour in both cases, however with varying speed of convergence.
Adapted figure from~\cite{Saga:2018nud}.}
\label{fig:Saga}
\end{figure}

The depicted collapse problem with initial data~\eqref{eq:Q1Dics} could also be solved directly with the recursion relations for the generic case (equations~\ref{eqs:LPT}), however then no analytical proofs of convergence are available. Instead, it turns out to be convenient to employ simple multi-scaling techniques, which in the present case amounts to impose a refined {\it Ansatz} for the components of the displacement $\fett{\xi}=\fett{x}-\fett{q}$ \cite{Rampf:2017jan}
\be \label{eq:quasi1D}
  \xi_i =  \delta^1_i F(q_1,\tau) + \epsilon \,\psi_i(\fett{q},\tau)  + O(\epsilon^2) \,,
\ee
\end{subequations}
where $F$ is the displacement part of the one-dimensional problem, while $\epsilon\, \fett{\psi}$ is a leading-order correction to the displacement. Supplemented with the boundary conditions~\eqref{eq:slaving}, 
it is straightforward to solve the master equations~\eqref{eqs:masterSingle} to successive powers in $\epsilon$. For example, at order $\epsilon^0$, one recovers the Zel'dovich solution $F= -\tau \sin q_1$, while at order $\epsilon^1$, one finds refined recursion relations for the Taylor coefficients appearing in $\fett{\psi}(\fett{q},\tau)=\sum_{n=1}^\infty \fett{\psi}^{(n)}(\fett{q}) \tau^n$ (see equation 36 in \cite{Rampf:2017jan}).
Most importantly, the refined recursion relation for $\fett{\psi}^{(n)}$ can be used to investigate, by analytical means, the asymptotics of the Taylor series at large orders, from which it follows that $\fett{\xi}$ is an entire function of time $\tau$. Thus, the quasi-one-dimensional problem is analytically solvable to arbitrary high precision.
See Fig.\,\ref{fig:Saga} for comparisons of various analytical solutions against numerical simulations (black dotted line): the blue (dot-dashed) line denotes the result using the {\it Ansatz}~\eqref{eq:quasi1D} up to second order in~$\epsilon$ (derived in~\cite{Saga:2018nud}), while the red-dashed line is obtained using~\eqref{eqs:LPT} up to the 10th order in the series. 
The left panel in Fig.\,\ref{fig:Saga} denotes the stated quasi-one-dimensional problem, while the right panel shows the collapse for tri-axial sine-wave initial conditions (see~\cite{Saga:2018nud} for the specific set-up).
Note that due to the use of trigonometric functions in the initial data, the theoretical solutions for the displacement are combinations of trigonometric functions as well -- thus,  no numerics are required for the shown theoretical solutions.

\paragraph{Quasi-spherical collapse.} There exists an exact parametric solution to the collapse of a  homogeneous, spherically symmetric over-density (sometimes called the top-hat model) \cite{Peebles1967}. This solution is however not obtained from the Vlasov--Poisson equations, but instead by a special solution of the Friedmann equations. Nonetheless, it can be shown that this simplified model can also be realised within the context of a Cartesian formulation of the Vlasov--Poisson equations \cite{Bernardeau:1992zw,Munshi:1994zb,Tatekawa:2004mq}.

By applying equivalent multi-scaling techniques in Lagrangian coordinates as reviewed above, it has been shown that the matter collapse of arbitrary small departures from spherical symmetry also constitutes an exact solution of~\eqref{eqs:masterSingle} until shell-crossing~\cite{Rampf:2017tne}. In that case, the initial gravitational potential can be taken to be
\begin{subequations}
\begin{align}
  \varphi^{\rm ini} &= K |\fett{q}|^2/6 + \epsilon \phi^{\rm ini}(q_1,q_2,q_3) \,,
\intertext{where $K$ is a positive [negative] constant amplitude of a spherical over-density [under-density/void], and $\phi^{\rm ini}$ is an arbitrary function that reflects the departure from spherical symmetry. The mathematical form of $\varphi^{\rm ini}$ suggests the solution {\it Ansatz}}
  \xi_i &= S(\tau) \, q_i + \epsilon \,\psi_i(\fett{q},\tau) \,,
\end{align}
\end{subequations}
where $S$ is a purely time-dependent function that models the purely spherical collapse, while $\epsilon\,\fett{\psi}$ reflects a perturbative departure therefrom. 

In the present case, both $S$ and $\fett{\psi}$ can be represented individually by convergent time-Taylor series~\cite{Rampf:2017tne}. However, the speed of convergence is significantly slower as compared to the quasi-one-dimensional case (cf.\ the LPT predictions in left versus right panel of Fig.\,\ref{fig:Saga}). The reason for the slow convergence is the presence of nearby singularities at an $O(\epsilon)$ amount of time after  shell-crossing. Actually, 
in the spherical case with $\epsilon \to 0$, the speed of convergence is even worse and furthermore comes with a singular velocity at shell-crossing (see also Fig.~1 in~\cite{Saga:2018nud} for the highly related triaxial symmetric case, or~\cite{Nadkarni-Ghosh:2010she} for a complementary analysis with similar conclusions). Nonetheless, we should stress that such singularities are basically man-made: a spherical top-hat over-density is a vastly simplified collapse model that has zero probability to occur in a Universe with random initial conditions~\cite{Doroshkevich1970}.

\subsection{Shell-crossing solutions for cosmological initial conditions} \label{sec:sc-random}

Investigating solutions until shell-crossings for random initial conditions is in general only feasible by semi-analytical avenues. This is so, since the Fourier transforms needed to solve for the Helmholtz decomposition~\eqref{eq:Helm} cannot be performed explicitly. Instead one typically resorts to Fast Fourier transforms, thereby determining the Taylor coefficients of~\eqref{eqs:LPT} on a three-dimensional numerical grid with $N$ collocation points.

However, there are also ways to  prove, by entirely theoretical means, that the solutions~\eqref{eqs:LPT} are time-analytic at least until some finite time, which goes back to a seminal paper by Zheligovsky \& Frisch \cite{Zheligovsky:2013eca}. In the following we sketch the essence of a normed-space proof, from which a lower bound on the radius of convergence of the series is obtained. Then, in a subsequent paragraph, the actual radius of convergence is determined by employing numerical extrapolation techniques.

\paragraph{Time-analyticity of dark-matter displacements.}
Instead of proving the convergence of $\fett{\xi}=\sum_{n=1}^\infty \fett{\zeta}^{(n)}\,\tau^n$, it is more instructive to first prove the convergence of the tensorial gradient therefrom, i.e., $\nab\, \fett{\xi}=\sum_{n=1}^\infty \nab\, \fett{\zeta}^{(n)}\,\tau^n$, with Taylor coefficients 
\be
  \nab \fett{\zeta}^{(n)} = \nab  \,  \nabla^{-2} \left( \nab L^{(n)} - \nab \times \fett{T}^{(n)} \right) \,, \label{eq:HelmGrad}
\ee
where $L^{(n)}$ and $\fett{T}^{(n)}$ are given by the recursion relations~\eqref{eq:scalarRec} and~\eqref{eq:vectorRec}, respectively. Before proceeding let us briefly introduce the $\ell^1$ norm denoted with $\| \mathbf{f} \| := \sum_n |\mathbf{f}_{\fett{k}}| < \infty$, for any periodic tensor, vector or scalar function $\mathbf{f}(\fett{q}) = \sum_{\fett{k}} \mathbf{f}_{\fett{k}} \exp(\ii \fett{k}\cdot \fett{q})$, where $\mathbf{f}_{\fett{k}}$ are its Fourier coefficients. The corresponding function space has the algebra property $\| \mathbf{f}\, \mathbf{g} \| \leq \| \mathbf{f} \| \, \| \mathbf{g} \|$, and furthermore bounds the operator $\nab \nab \nabla^{-2}$ to unity from below. Using these properties as well as the explicit forms of $L^{(n)}$ and $\fett{T}^{(n)}$ (equations~\ref{eq:scalarRec} and~\ref{eq:vectorRec}), the $\ell^1$ norm of~\eqref{eq:HelmGrad} turns into the polynomial inequality \cite{Zheligovsky:2013eca}
\be \label{ineq1}
\!\!\!\! \| \nab \fett{\zeta}^{(n)} \| \leq  \| \nab \nab \varphi^{\rm ini} \|\, \delta_1^n + 12\!\!\! \sum_{i+j=n} \!\! \| \nab \,\fett{\zeta}^{(i)} \|   \| \nab \fett{\zeta}^{(j)} \|  + 6 \!\!\!\!\sum_{i+j+k=n} \!\! \| \nab \fett{\zeta}^{(i)} \|   \| \nab \fett{\zeta}^{(j)} \|   \| \nab \fett{\zeta}^{(k)} \| .
\ee
In deriving this inequality, one considers the large-$n$ limit (relevant for investigating convergence), where the coefficients appearing in the various terms in  $L^{(n)}$ and $\fett{T}^{(n)}$ are bounded (rational) functions. Now, introducing the generating function 
$
  \zeta(\tau) := \sum_{n=1}^\infty \| \nab \fett{\zeta}^{(n)} \| \tau^n
$,
as well as multiplying~\eqref{ineq1} with $\tau^n$ and summing over $n$ from one to infinity, one gets
$\zeta \leq  \| \nab \nab \varphi^{\rm ini} \| \tau + 12 \zeta^2 + 6\zeta^3$, which is equivalently
\be
  p(\tau) := \| \nab \nab \varphi^{\rm ini} \| \tau + 12 \zeta^2 + 6\zeta^3 - \zeta \geq 0 
\ee
(see \cite{Rampf:2015mza} for a graphical representation as well as for the $\Lambda$CDM case).
Initially for $\tau=0$, this polynomial $p$ behaves as $-\zeta$ for small $\zeta$ and asymptotically as~$\zeta^3$; thus, $p(\tau=0)$ has three roots with one at $\zeta_{\rm phys} =0$. For small positive times, $p$ gets shifted upwards, 
thereby drifting the root $\zeta_{\rm phys}$ to the right. This branch from $\zeta=0$ until $\zeta_{\rm phys}$
marks the physical regime where $p>0$ is bounded (i.e., the Taylor coefficients of $\zeta$ do not blow up).
However, at some critical time, denoted with $\tau_{\rm c}$, this boundness property will be lost as the asymptotic behaviour $\zeta^3$ kicks in. It is found that this critical time is achieved precisely when the root $\zeta_{\rm phys}$ merges with another root; therefore $\tau_{\rm c}$ can be determined by setting the discriminant of $p(\zeta)$ to zero, leading to \cite{Zheligovsky:2013eca}
\be \label{eq:bound}
  \tau_{\rm c} = \frac{c}{\| \nab \nab \varphi^{\rm ini} \|} \,, \qquad c = 3\sqrt{2} - 38/9 \simeq 0.0204 \,.
\ee
Thus, the displacement is guaranteed to be time-analytic between $0 \leq \tau \leq \tau_{\rm c}$, while 
the value of the lower bound $\tau_{\rm c}$ crucially depends on the Hessian of the initial gravitational potential. 
We note that the above bound can be slightly improved \cite{Zheligovsky:2013eca,Michaux:2020yis}, especially when assumptions about the regularity of $\varphi^{\rm ini}$ are imposed.

\paragraph{Shell-crossing and radius of convergence.}
To determine the actual radius of convergence of $\fett{\xi}(\fett{q},\tau) = \sum_{s=1}^\infty \fett{\zeta}^{(s)}(\fett{q}) \,\tau^s$ for random initial conditions, one has to resort to numerical tests or extrapolation methods.
One of such tests is to verify whether theoretical predictions, such as time and location of the first shell-crossing, converges for successively higher orders in the perturbative truncations $n$ of the displacement series. That is
one searches for the location where the perturbatively truncated Jacobian
\be \label{eq:Jn}
 J^{\{ n \}} := \det \Big( \nabq \Big[ \fett{q} + \sum_{s=1}^n \boldsymbol{\zeta}^{(s)}(\fett{q})\,\tau^s \Big] \Big) 
\ee
vanishes for the first time (cf.\ equation~\ref{eq:SCcondition}) \cite{Rampf:2020hqh}.
For each truncation order $n$, one then obtains perturbative estimates of the shell-crossing time and location,  denoted respectively with $\tau_\star^{(n)}$ and $\fett{q}_\star^{(n)}$. By monitoring the trend of these estimates at successively high orders $n$, one then obtains evidence for or against (weak) convergence.

In \cite{Rampf:2020hqh}, the recursive displacement solutions~\eqref{eqs:LPT} as well as $J^{\{ n \}}$ have been realised on a spatial grid with $N$ collocation points for cosmological random initial conditions. The corresponding code\footnote{Available from \url{https://bitbucket.org/ohahn/monofonic}.} is parallelised (MPI+threads), and de-aliases cubic functions in the displacement. Most analysis in~\cite{Rampf:2020hqh} was performed with a grid resolution of $N= 256^3$, for which convergence of the shell-crossing time $\tau_\star$ accurate to three significant digits was typically reached between orders $n=6-10$.

\begin{figure}
  \begin{center}
  
  \includegraphics[width=\textwidth]{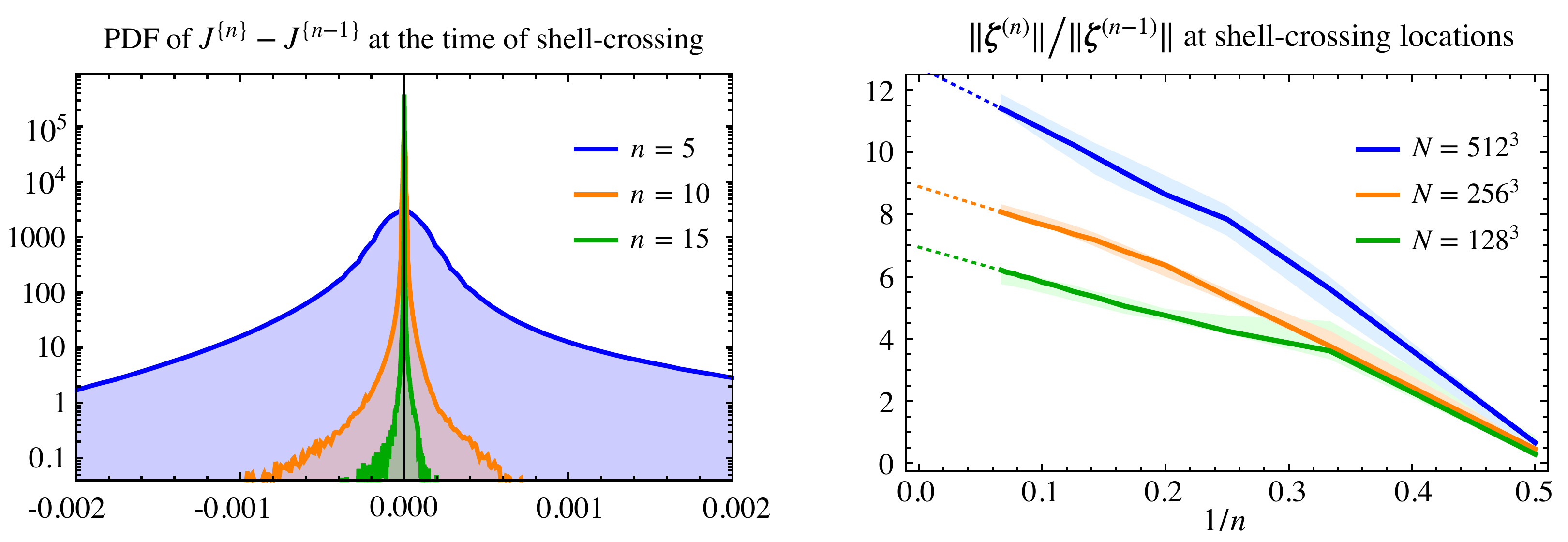}

  \end{center}
\caption{Convergence studies of~\eqref{eqs:LPT} for cosmological random initial conditions; figure adapted from~\cite{Rampf:2020hqh}. {\bf Left panel:} One-point distribution of the residual $J^{\{n\}}-J^{\{n-1\}}$ at the orders $n=5,10,15$ (shown in blue, orange and green, top to bottom), evaluated at $256^3$ Lagrangian collocation points at the time of first shell-crossing. This figure demonstrates the qualitative convergent behaviour in the whole spatial domain.
{\bf Right panel:} Domb-Sykes plot for the ratio $\| \fett{\zeta}^{(n)}\|/\| \fett{\zeta}^{(n-1)}\|$ for $1 \leq n \leq 15$, evaluated at the Lagrangian location that, for given random realisation, shell-crosses first. 
Shown results employ a spatial grid resolution of  $N=512^3, 256^3, 128^3$  (shown in blue, orange and green, top to bottom).
The solid lines denote the median while the shaded regions are the 32 and 64 percentiles obtained from five random realisations. }
\label{fig:convergenceLPT}
\end{figure}

As another convergence test, we show in the left panel of Fig.\,\ref{fig:convergenceLPT} the one-point probability density function (PDF) of the residual $\Delta J^{\{ n \}} := J^{\{ n \}} - J^{\{ n -1\}}$ from all Lagrangian grid points, evaluated at the time of first shell-crossing (here at $\tau_\star = 0.0813$ for a standard $\Lambda$CDM cosmology with $N=256^3$). Evidently, $\Delta J^{\{ n \}}$ peaks sharper at the origin for higher orders, indicating that truncation errors decrease rapidly at increasing order~$n$, which is an expected feature of a convergent series.

Next we determine numerically the radius of convergence, for which it is useful to consider the closely related series of the $\ell^2$ norms of displacement coefficients, i.e.,
\be
  \| \fett{\xi}(\fett{q},\tau) \| := \sum_{n=1}^\infty \big\|  \fett{\zeta}^{(n)}(\fett{q}) \big\| \,\tau^n  \,. 
\ee
To investigate its convergence, one can perform d'Alembert's ratio test which states
\be \label{eq:ratio}
  \frac{1}{\tau_R} = \lim_{n\to \infty} \frac{\| \fett{\zeta}^{(n)} \|}{\| \fett{\zeta}^{(n-1)} \|}   \,,
\ee
where $\tau_R$ is the radius of convergence -- when that limit exists.
A standard way to estimate $\tau_R$ is the Domb--Sykes plot \cite{DombSykes1957}, where one draws $\| \fett{\zeta}^{(n)} \| / \| \fett{\zeta}^{(n-1)} \|$ versus $1/n$, from which $1/\tau_R$ follows by extrapolation to the $y$-intercept (see also \cite{Michaux:2020yis,Podvigina++2016} for related applications).
In the right panel of Fig.\,\ref{fig:convergenceLPT} we show the Domb-Sykes plot for the spatial coefficients $\|\fett{\zeta}^{(n)}\|$ evaluated at the Lagrangian locations of first shell-crossing for three different resolutions, $N=512^3, 256^3, 128^3$ (respectively in blue, orange and green; top to bottom), while the length of the grid is fixed at $L = 125\,$Mpc$h^{-1} \simeq 5.51 \times 10^{24}\,$m.
The shaded areas denote its 32 and 68 percentiles obtained from~5~random realisations, while the solid lines denote its median.
For sufficiently large Taylor orders ($n\gtrsim 7$) the ratios settle into a linear behaviour, thereby justifying a linear extrapolation to the $y$-intercepts (dotted lines), which, using~\eqref{eq:ratio}, leads to the radius of convergence~$\tau_R$. In all cases considered the (temporal) value of~$\tau_R$ is roughly~$30-40$\% larger than the shell-crossing time (see table~1 in~\cite{Rampf:2020hqh} for specific values), thus indicating that the mathematical validity of~\eqref{eqs:LPT} surpasses vastly its physical validity (since~\ref{eqs:LPT} is only valid until shell-crossing). This is an understood phenomenon, for example the one-dimensional and quasi-one-dimensional shell-crossing solutions have an infinite radius of convergence (see section~\ref{sec:sc-anal}).

Obviously, $\tau_R$ depends on the chosen resolution: since we keep the grid length fixed, this mostly reflects the known dependency of theoretical solutions on UV physics beyond the Nyquist frequency; see e.g.\ \cite{Bernardeau:2001qr,Michaux:2020yis,Taruya:2018jtk}; see however \cite{Schmidt:2020ovm} for a recent investigation of   residual aliasing effects.

Finally, the linear behaviour of the ratios of coefficients at large orders suggests that the convergence limiting singularities have a local behaviour of \cite{Rampf:2020hqh}
\be
 \| \fett{\xi}(\fett{q},\tau) \| \propto (\tau - \tau_R)^{\,\alpha} \,,
\ee
where $\alpha$ is a singularity exponent. In this case, the ratio of coefficients satisfies the linear relationship (cf.\ \cite{Podvigina++2016,vanDyke1974})
\be
   \frac{\| \fett{\zeta}^{(n)} \|}{\| \fett{\zeta}^{(n-1)} \|} = \frac{1}{\tau_R} \left[ 1- (1+\alpha) \frac 1 n \right] \,.
\ee
Thus, the slope of the linear extrapolations reveal the singularity exponents, which in the considered cases are $\alpha = 0.61, 0.38, 0.56$ respectively for the resolutions $N= 128^3, 256^3, 512^3$ \cite{Rampf:2020hqh}. Since these exponents are positive non-integers and smaller than unity, it follows that the first time derivative of $\| \fett{\xi}(\fett{q},\tau) \|$  blows up at $\tau=\tau_R$.

Concluding this section, the recursive solutions~\eqref{eqs:LPT} for $\fett{\xi} =\sum_{n=1}^\infty \fett{\zeta}^{(n)} \tau^n$ appear to deliver physically and mathematically meaningful results until the time of shell-crossing $\tau_\star$, while the various bounds and limits are related through $0 < \tau_{\rm c}  < \tau_\star < \tau_R$, 
where $\tau_{\rm c}$ is an entirely analytical estimate of the radius of convergence, while $\tau_R$ is the actual radius of convergence determined through numerical extrapolation techniques.

%%%%%%%%%%%%%%%%%%%%%%%%%%%%%%%%%%%
\section{Numerical and analytical post-shell-crossing methods (\texorpdfstring{${\cal M} \neq 0$}{M!=0})} \label{sec:generalPSC}

Solving the Vlasov--Poisson equations beyond shell-crossing is a challenge that is usually attempted by numerical $N$-body simulations, where (very massive) macro-particles are employed to coarse sample the dark-matter distribution in phase-space (see e.g.\ \cite{BertschingerGelb,Dehnen:2011fj} for reviews). However, these simulations can be prone to errors introduced through the discrete representation of the phase-space distribution (e.g.~\cite{Abel:2011ui,Melott:1996wz,Angulo:2013qp,Colombi:2014zga}).

Recently, however, there have been  encouraging avenues at various fronts using elementary methods as well as (semi-)analytical and numerical techniques. While this review predominantly focuses on the progress related to elementary methods for Vlasov--Poisson (section~\ref{sec:PSC}), in the following two sections, we attempt to summarise important findings related to semi-analytical/effective avenues (section~\ref{sec:effective}) as well as to novel simulation techniques (section~\ref{sec:num}).

\subsection{Semi-analytical/effective methods}\label{sec:effective}

Simple (time-)Taylor expansions, such as~\eqref{eqs:LPT} in Lagrangian coordinates, cease to be valid at the time of first shell-crossing \cite{Pietroni:2018ebj,Rampf:2019nvl}. Of course, the same can be said when seeking theoretical solutions to Vlasov--Poisson in Eulerian coordinates $\fett{x}$, where the density and velocity can be Taylor expanded until shell-crossing with equivalent methods as reviewed in section~\ref{sec:strategy} (dubbed SPT; see also e.g.~\cite{Bernardeau:2001qr,Buchert:1997dr}).

However, there also exist  methods that seek to incorporate multi-streaming effects in a rather effective way \cite{Buchert:2005xj,Pueblas:2008uv}, notably the so-called coarse-grained perturbation theory \cite{Pietroni:2011iz} as well as the effective theory of large-scale structure \cite{Baumann:2010tm,Carrasco:2012cv,Porto:2013qua}. These approaches have a few properties in common, in particular that they are typically not designed to solve for the dark-matter evolution at the deterministic level; instead predictions  are of statistical nature (e.g., for the power spectrum of the matter density).

Another important characteristic of effective approaches is that they attempt to circumvent the obvious shortcomings by first ``smoothing out'' the small-scale physics -- which can be strongly influenced by shell-crossing/multi-streaming dynamics, beyond a spatial cut-off scale~$1/\Lambda$. Subsequently, the large-scale physics is handled with Taylor/perturbative expansions, while the small-scale/multi-streaming physics is incorporated through certain outputs from numerical simulations (e.g., velocity dispersion),  or even through cosmological observations \cite{DAmico:2019fhj,Colas:2019ret}. 

At the technical level, this can be achieved  by introducing a spatially smoothed (large-scale) distribution function
$f_\Lambda(\fett{x},\fett{p},t) := \int  W_\Lambda(\fett{x}-\fett{x}') f(\fett{x},\fett{p},t) \dd^3 x'$, where
$f$ is the distribution function that solves equation~\eqref{eq:VP}, while $W_\Lambda$ is a Window function which is usually taken to be a normalised Gaussian, i.e., $W_\Lambda \propto \exp(-\Lambda^2 |\fett{x}|^2/2)$. Then, 
by taking kinetic moments of the resulting large-scale Vlasov--Poisson equations, one obtains fluid-type equations for an effective large-scale fluid  \cite{Baumann:2010tm,Carrasco:2012cv}
\be \label{eq:effective}
 \!\!\!\!\!\partial_t \delta_\Lambda + \nab \cdot [(1+\delta_\Lambda) \fett{u}_\Lambda] =0 , \quad \,\,\,  
  \tfrac{\dd}{\dd t} \fett{u}_\Lambda + \text{\small $2$} \tfrac{\partial_t a}{a} \fett{u}_\Lambda  = \tfrac{1}{a^2} \Big( - \nab \varphi_\Lambda 
  - \tfrac{1}{\rho_\Lambda}\nab \cdot [\rho_\Lambda \fett{\sigma}]_\Lambda \Big) ,
\ee
where $\rho$, $\rho \fett{u}$ and $\fett{\sigma}$ are respectively the zeroth, first and second kinetic moments of the distribution function (see e.g.\ \cite{Bernardeau:2001qr}), while $[\mathbf{X}(\fett{x})]_\Lambda :=  \int  W_\Lambda(\fett{x}-\fett{x}')\mathbf{X}(\fett{x}')\, \dd^3 x'$ for any scalar, vector or tensor function $\mathbf{X}(\fett{x})$.

Equations~\eqref{eq:effective} state respectively mass and momentum conservation of a fluid that is subject to a source term due to small-scale physics, here introduced through the velocity dispersion tensor $\fett{\sigma}$.
Note that $\fett{\sigma}$ generally also includes a non-zero pressure contribution which physically arises due to multi-streaming. How precisely $\fett{\sigma}$ is incorporated, depends on specific model details for which we kindly refer to the original references.  

Finally, an entirely different, semi-analytical approach is the so-called kinetic field theory, which 
is a microscopic statistical field theory that is intimately linked to the classical Hamiltonian approach (e.g.~\cite{Bartelmann:2014gma,Lilow:2018ejs,Bartelmann:2019unp}). The central building block is the generating functional involving the $N$-particle action, which physically encodes the classical transition probability from the initial and final field state. 
To incorporate particle interactions (relevant for multi-streaming), perturbative expansions and/or approximations are necessary. Furthermore, applying suitable limits, particle shot-noise is removed and, therefore, the theory should deliver meaningful approximations to statistical predictions of Vlasov--Poisson. So far, the corresponding methods have been only worked out at leading order, therefore statements about convergence can not be made at this stage.

\subsection{Numerical simulations techniques}\label{sec:num}

Standard $N$-body methods solve the Hamiltonian equations of  motion~\eqref{eqs:HamiltonEoMs} for a set of $N$ macro-particles, where the continuous phase-space distribution is effectively coarse-sampled. 
There are various types of $N$-body simulations that essentially differ in the way how the Poisson equation is solved, e.g., using standard particle-in-cell / particle mesh \cite{Doroshkevich1980,Hockney1981} or brute force methods \cite{Aarseth63,Aarseth79}, codes with adaptive mesh refinements \cite{Villumsen,Suisalu:1994jj,Kravtsov:1997vm,Knebe:2001av,Teyssier:2001cp,ENZO:2013hhu}, tree algorithms \cite{Barnes:1986nb,Hernquist1987,Springel:2000yr} and, finally, combinations of various algorithms \cite{Efstathiou:1985re,Hockney1981,Couchman1991,Bagla:1999tx,Springel:2005mi,Garrison:2017ssz}; see e.g.~\cite{Dolag:2008ki,Vogelsberger:2019ynw,AnguloHahn2021} for extensive reviews about $N$-body methods.

In the following, we provide an overview of recent numerical approaches that aim to retrieve fine-grained details of the dark-matter phase-space, and thus going, in one way or another, beyond  standard $N$-body techniques.

\paragraph{Phase-space reconstruction.} 
With the standard  $N$-body technique, quasi-continuous estimates of observables, such as the matter density or (mean) velocity,  can be obtained by applying suitable smoothing operations on the set of discrete tracer particles (e.g., the cloud-in-cell interpolation that assigns a density cloud around particles).
Recently, however, there have been novel approaches that are able to reconstruct the dark-matter phase-space to much higher accuracy, while at the heart still relying on the $N$-body technique.

One of these new approaches is dubbed GDE which is short for geodesic deviation equation \cite{Vogelsberger:2008qb,Vogelsberger:2010gd}. There, on top of the standard equations of motion, one solves for the evolution of the tangent space in the 6D phase-space around each tracer particle, thereby locally recovering some details of the dark-matter sheet. This approach has turned out to be fruitful at the very final stages during halo formation, where the number of streams can easily go into the several hundreds. On the other hand, the approach can suffer artificial fragmentation in certain (warm) dark-matter scenarios, which arises during the rather early gravitational evolution~\cite{Wang:2007he,Melott:2007fq}.

Another approach is the sheet method, which employs interpolation/tessellation techniques to recover the dark-matter phase-space sheet from the set of discrete tracer particles \cite{Hahn:2015sia} (see also \cite{Abel:2011ui,Shandarin:2011jv,Hahn:2012ma}).  The sheet reconstruction (with isotropic mesh refinements) provides continuous density estimates and thus an improved computation of  particle forces during the subsequent stages of the gravitational evolution. This also alleviates the aforementioned problem of artificial fragmentation.
However, the dark-matter sheet reconstruction fails at the very final stages, especially inside halos, due to the increased complexity of phase-space dynamics. 
Recently, the sheet and GDE approaches have been married, thereby allowing to efficiently exploit the strengths of both methods~\cite{Stucker:2019txm}.

\paragraph{Pure phase-space tessellation methods.} Instead using tracer particles,
the numerical code 
{\sc ColDICE} employs simplices (tetrahedra) to tessellate the continuous dark-matter sheet \cite{Sousbie:2015uja}.
Fairly similarly as in the above sheet method, the forces are obtained from continuous density estimates and subsequent interpolations to an undistorted grid.  
Furthermore,
{\sc ColDICE} allows for anisotropic mesh refinements, thereby being able to resolve the phase space  
to high precision. Unfortunately, the complexity of the dark-matter sheet becomes computationally too demanding at the late stages during halo formation.
Nonetheless, important lessons can be drawn from using such advanced approaches to Vlasov--Poisson; see in particular \cite{Colombi:2020xbv} for explicit comparisons of {\sc ColDICE} against standard $N$-body codes during the early violent relaxation phase of halo formation
(see also~\cite{Colombi:2014zga}).

\paragraph{Vlasov--Poisson simulations in 6D.} There exist also full Vlasov--Poisson simulations that are not constrained in evolving the dark-matter sheet \cite{Yoshikawa:2012yj,Tanaka_2017}. Such avenues are especially important when e.g.~investigating the phase-space in the presence of warm components (e.g., neutrinos, or warm dark matter) \cite{Yoshikawa:2020ehd,Colombi:2017qww}.

\paragraph{Vlasov--Poisson through Schr\"odinger--Poisson.}
Lastly, an effective Vlasov simulation is realised by exploiting the correspondence to the quantum mechanical
Schr\"odinger--Poisson equations \cite{Widrow:1993qq}. Indeed in \cite{Mauser2002} it has been proven for the 1D case that this correspondence is exact in the limit $\hbar\to0$. By contrast, keeping $\hbar$ small but finite in Schr\"odinger--Poisson delivers meaningful approximations, where $\hbar$ acts now as an effective coarse-graining scale in the phase-space; see e.g.~\cite{Mocz:2017wlg,Kopp:2017hbb,Garny:2019noq,Eberhardt:2020otp}.

\subsection{Mathematical post-shell-crossing analysis}\label{sec:PSC}

Turning back to the theoretical solution scheme for the Lagrangian master equations~\eqref{eqs:formalPSCsolution}, the leading-order behaviour of the involved source terms shortly after shell-crossing time $\tau_\star$ is given by
${\cal W}(\fett{x}(\fett{q},\tau)) \simeq {\cal W}(\fett{x}_{\!\fett{\bullet}}(\fett{q},\tau))$ and ${\cal M}(\fett{x}(\fett{q},\tau)) \simeq {\cal M}(\fett{x}_{\!\fett{\bullet}}(\fett{q},\tau))$,
which is, strictly speaking, exact at $\tau =\tau_\star$, and approximatively thereafter, since $\fett{x}_{\!\fett{\bullet}}(\fett{q},\tau) = \fett{q}+ \sum_{n=1}^{n_{\rm max}} \fett{\zeta}^{(n)}(\fett{q})\,\tau^n$ constitutes only a solution of Vlasov--Poisson until $\tau_\star$.

Let us show how ${\cal M}$ can be determined, which can also be written as
\begin{align}
 {\cal M}(\fett{x}_{\!\fett{\bullet}}(\fett{q},\tau)) &=1    - \det[\nabq \fett{x}_{\!\fett{\bullet}}(\fett{q},\tau)]  \int \delta_{\rm D} \left[ \fett{x}_{\!\fett{\bullet}}(\fett{q},\tau)- \fett{x}_{\!\fett{\bullet}}(\fett{q}',\tau) \right] \, \dd^3 q' \nonumber \\
   &= 1 - \nabq \cdot \int \vec{\Theta} \left[ \fett{x}_{\!\fett{\bullet}}(\fett{q},\tau)- \fett{x}_{\!\fett{\bullet}}(\fett{q}',\tau) \right] \, \dd^3 q' \,, \label{eq:Heaviside}
\end{align}
with $\nabq \cdot \vec{\Theta}(\fett{q}) := \delta_{\rm D}^{(3)}(\fett{q})$, where $\vec{\Theta}$ would correspond to the standard Heaviside  function in one space dimensions, while in 3D, it may be considered as a gravito-electric field. 
Although not strictly necessary, for the proceeding calculations it turns out to be convenient to employ $\vec{\Theta}$, although similar (but substantially longer) arguments hold when using the Dirac delta instead. 
Evaluating the integral in~\eqref{eq:Heaviside}  boils down to determining the length of the branches when the argument of $\vec{\Theta}$ is positive. 
Obviously, for this one needs to determine the roots $\fett{q}'$ of 
\be
  \fett{x}_{\!\fett{\bullet}}(\fett{q},\tau)- \fett{x}_{\!\fett{\bullet}}(\fett{q}',\tau) \stackrel ! = \fett{0}\,,
\ee 
which is a transcendental equation for which exact solutions are generally not available.
To proceed, one considers normal forms, which are spatial Taylor expansions of the map $\fett{x}_{\!\fett{\bullet}}(\fett{q},\tau)$ around the location $\fett{q}_\star$ of first shell-crossing, combined with exploiting the expected topology of $\fett{x}_{\!\fett{\bullet}}(\fett{q},\tau)$ for times shortly after shell-crossing.

What do we mean by the last statement? Consider the left panel in Fig.\,\ref{fig:map-density}, which shows the map in 1D: at shell-crossing time (here $\tau_\star =1$ at $q_\star =0$), the map has an inflection point with local behaviour $\sim q^3$ around the shell-crossing location; shortly later (here shown at $\tau =1.3$ in blue) the local behaviour of the map in 1D is \cite{Rampf:2019nvl,Colombi:2014lda,Taruya:2017ohk}
\be \label{eq:normal}
  x_{\!\fett{\bullet}}(q,\tau) \simeq q + \tau \zeta^{\rm N}(q)\,, \qquad \zeta^{\rm N}(q) = \left[ - c_1 (q-q_\star) + c_3 (q-q_\star)^3 \right] \,,
\ee
where $c_{1,3}$ are  positive Taylor coefficients. Here,  $\zeta^{\rm N}(q)$ corresponds to the said normal form in the one-dimensional case, and note specifically that a term $\sim (q-q_\star)^2$ is absent (it can be removed by a suitable Galilean transformation). Employing~\eqref{eq:normal} to determine the roots of
$x_{\!\fett{\bullet}}(q,\tau)- x_{\!\fett{\bullet}}(q',\tau) = 0$ is straightforward, and in fact a good approximation for times shortly after the first shell-crossing (at later times higher refinements become necessary; see \cite{Rampf:2019nvl,Colombi:2014lda,Taruya:2017ohk}).
Generalisations of the above to the three-dimensional case are in principle straightforward yet tedious and, so far, have not been reported in the literature.
Once  ${\cal M}(\fett{x}_{\!\fett{\bullet}}(\fett{q},\tau))$  
is determined, the post-shell-crossing displacement is readily obtained using equations~\eqref{eqs:formalPSCsolution}; see e.g.\ equation~(18) in~\cite{Rampf:2019nvl} for explicit solutions in the 1D case.

\begin{figure}
  \begin{center}
  
  \includegraphics[width=0.95\textwidth]{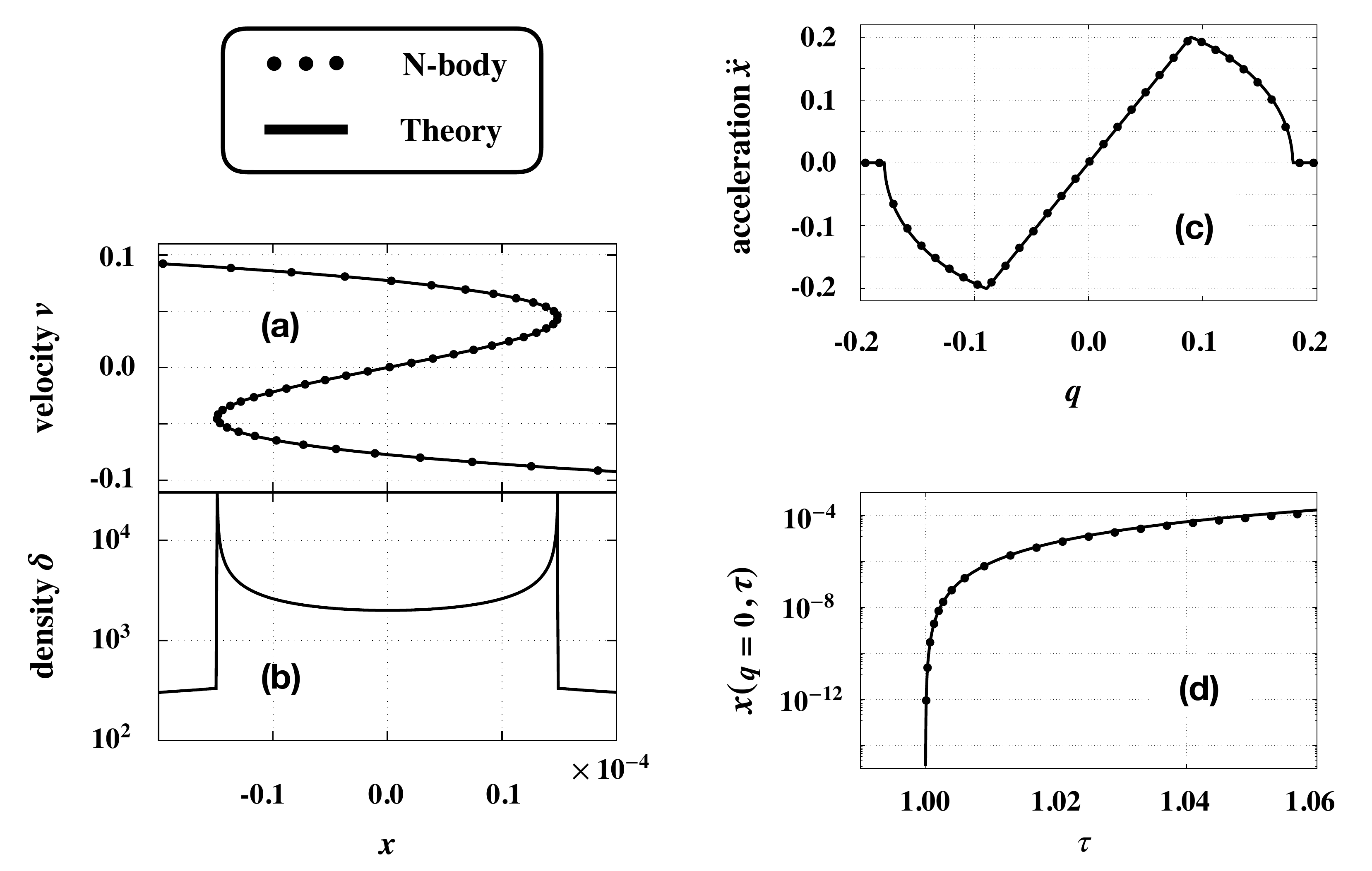}

  \end{center}
\caption{Comparison of fully analytical predictions (solid lines) against numerical simulations (dotted lines) after the first shell-crossing, for the same one-dimensional setup as for Fig.\,\ref{fig:phasespace-color} on the torus with $x,q \in [-\pi,\pi)$. The panels (a)--(c) are evaluated at $\tau=1.001$ (shell-crossing at $\tau=1$), and show respectively the dark-matter phase-space, the projected density contrast, and the particle acceleration. 
Panel~(d) shows the temporal evolution of the particle at  $q=0$ (i.e., the shell-crossing location), which suddenly begins moving at shell-crossing time $\tau=1$ due to an asymmetry in the initial data. Panels~(c) and~(d) display non-analytic behaviour where derivatives in the phase-space blow up. Figures obtained using the methods of~\cite{Rampf:2019nvl}.
}
\label{fig:PSC}
\end{figure}

These purely analytical solutions are shown in Fig.\,\ref{fig:PSC} and compared against numerical simulations at high resolution (see footnote~\ref{footnote1Dsim}). Panel (a) shows the phase-space in one dimension
at time $\tau=\tau_1 =1.001$ featuring single- and multi-streaming regions that are spatially separated by infinite densities (panel b). At the same time, panel (c) shows the corresponding acceleration of particles, displaying four non-differentiable sharp features~\cite{Rampf:2019nvl}: in single-stream regions the acceleration is exactly zero (very left and very right regions in panel c), but, at the depicted time, then jumps to non-zero values with local behaviour $\xi \sim (\tau - 1.001)^{5/2}$ and $\xi \sim (q \pm 0.17)^{5/2}$ at time $\tau_1$, thereby indicating that the third derivatives in space and time of the displacement blow up.
Furthermore, the third space derivative of the displacement flips sign (here around $q = q_2 =\pm 0.08$)
thereby marking a singularity of $\xi \sim (q-q_2)^3 \Theta(q_2-q)$ and similarly in the temporal dependence.

Finally, a non-trivial singularity appears when the initial velocity is not exactly point-symmetric, here for a simplified model with $v^{\rm ini} = - \sin q + 0.1 \sin^4 q - 0.12 \sin^6 q$: the particle that is initially at $q=0$ will remain there until shell-crossing ($\tau=1$), at which time an asymmetric multi-streaming force arises, essentially as a consequence of momentum conservation, that leads the $q=0$ particle to begin moving with $\xi(q=0) \sim \delta \tau^3$~\cite{Rampf:2019nvl} (panel d in Fig.\,\ref{fig:PSC}; see also~\cite{Pietroni:2018ebj} for similar conclusions in a slightly different set-up).
Of course, the above initial data just serves as a simplified model to investigate realistic collapse scenarios: there, the random nature of initial perturbations will always distort this point-symmetry and thus, the appearance of the presently described singularity is a generic phenomenon, expected for realistic (cosmological) initial conditions. We remark that to determine this highly non-trivial feature depicted in panel (d) of Fig.\,\ref{fig:PSC}, the invariance~\eqref{eq:Heck} of Vlasov--Poisson under non-Galilean transformations can be used.

Concluding this section, standard Taylor expansions of the displacement cease to be valid after shell-crossing due to the emergence of non-analytic behaviour in the phase-space, however by now there are promising avenues that allow to enter into the post-shell-crossing regime while unveiling  intrinsic properties of Vlasov--Poisson.

%%%%%%%%%%%%%%%%%%%%%%%%%
\section{Related problems in plasma physics}\label{sec:plasma}

Vlasov--Poisson is also relevant in plasma physics, in particular when a collisionless, electrically charged medium is exposed to electrostatic interactions. In what follows we discuss various plasma problems and lay out connections to the cosmological case.

\paragraph{One component plasma model (OCP).} Possibly one of the simplest plasma models consists of a set of $N$ electrons with charge $e$ and mass $m$ that interact only through electrostatic forces. In order to warrant charge neutrality, these electrons are immersed in a rigid uniform background of opposite charge, which in the present case is $-\bar \rho = - N e/V$ where $V$ is the spatial volume under consideration. 
Possible applications of the OCP includes the (rough) modelling of the interior of stars, in particular of white dwarfs (see e.g.~\cite{Baus++review1980,KoesterWhiteDwarfs}). Adopting somewhat the above notation, the equation of motion for the $j$th electron, formulated in physical coordinates in 3D and employing the standard time $t$,  reads  (see e.g.~\cite{Baus++review1980,Escande2018})
\be \label{eq:OCP}
 \ddot{\fett{x}}_j  = \frac e m \nabx \varphi(\fett{x}_j) , \qquad \nabx^2 \varphi = 4\pi (\rho - \bar\rho) \,, \qquad  \rho =   e \sum_{i=1}^N \delta_{\rm D}^{(3)}(\fett{x} - \fett{x}_i(t))  \,,
\ee
where $\nabx \varphi = \fett{E}$ corresponds to the electrostatic field, and
we note that the vacuum permittivity is $\varepsilon_0 = 1/(4\pi)$ in Planck units. By applying similar arguments as in the cosmological case, one may introduce a continuum description for the OCP by employing the Lagrangian map $\fett{q} \mapsto \fett{x}(\fett{q},t)$ from initial position $\fett{q}$ to the current Eulerian position $\fett{x}$. 
The plasma velocity can be represented as $\fett{v}(\fett{x}(\fett{q},t)) = \dot {\fett{x}}(\fett{q},t)$, where the over-dot denotes now the convective time derivative with respect to standard time, i.e., $\dd / \dd t = \partial / \partial_t |_{\fett{q}} = \partial/\partial t |_{\fett{x}}  + (\dd \fett{x}/\dd t)  \cdot \nabx$. 
Then, equations~\eqref{eq:OCP} may be written as a continuous Vlasov--Poisson description
\be \label{eq:VPsinglebeam}
  \ddot {\fett{x}}(\fett{q},t) = \frac e m \nabx \varphi(\fett{x}(\fett{q},t)),  \quad \nabx^2 \varphi = 4\pi (\rho - \bar\rho) , \quad  \rho =   e \int \!\! \delta_{\rm D}^{(3)} \left[ \fett{x}(\fett{q},t) - \fett x (\fett{q}',t) \right] \dd q'  .
\ee
In the following we discuss the similarities and differences of these equations as compared to their cosmic counterparts.  
While the former makes use of the standard time, the cosmic Vlasov--Poisson equations~\eqref{eq:VPLag} are formulated in the scale-factor time $\tau=a$ (since in the cosmic case, time-analyticity holds in~$\tau \sim t^{2/3}$ but not in~$t$). 
Also, equations~\eqref{eq:VPLag} feature a Hubble drag term $\sim \dot {\fett{x}}$ stemming from the present choice of spatio-temporal coordinates (note that the Hubble drag term formally disappears when using a new time variable $\eta$ defined through $\dd t= a^2 \dd \eta$ \cite{Buchert:1989xx,Shandarin1980}). 
Furthermore, equations~\eqref{eq:VPLag} have a minus sign in front of $\nabx \varphi$ while in~\eqref{eq:VPsinglebeam} there is a plus sign; 
this flip in sign just reflects the change of charge in the gravitational attractive and electrically repulsive case. 
Finally, observe that equations~\eqref{eq:VPsinglebeam} are invariant under the non-Galilean transformation $\fett{x} \to \fett{x} + \fett{n}(t)$, where $\fett{n}(t)$
is an arbitrary function of time; this invariance is in close resemblance with the one of the cosmological case that we discussed around equation~\eqref{eq:Heck}.

Perturbative techniques applied to variants of equations~\eqref{eq:VPsinglebeam} are frequently encountered  in plasma physics (e.g.~\cite{DiamondBook,Escande2016}), however, we are unaware of explicit use of Lagrangian coordinates (though they are implicitly assumed for equation~\ref{eq:OCP} in the discretised sense). In this context note that for a single electron particle at current position $\fett{x}(t)$ exposed to its own electric field, the Poisson potential is $\tilde \varphi(\fett{k}) = - (4\pi e /k^2)\, \exp[-\ii \fett{k} \cdot \fett{x}(t)]$ in Fourier space for $\fett{k}\neq0$ (cf.\ equation~1 in~\cite{Escande2016}). By contrast, for a set of electrons parametrised in Lagrangian coordinates $\fett{q}$, the Fourier transform of the Poisson potential is easily obtained from~\eqref{eq:VPsinglebeam}; it reads $\tilde \varphi(\fett{k}) = - (4\pi e /k^2) \int \exp[-\ii \fett{k} \cdot \fett{x}(\fett{q},t)]\,\dd^3 q$ for $\fett{k}\neq0$.

Finally, note that as long as the plasma is monokinetic (single-beam), equations~\eqref{eq:VPsinglebeam} can equivalently be formulated in Eulerian coordinates as a set of closed fluid equations
\be \label{eq:OCPeuler}
  \frac{\partial }{\partial t} \fett{v} + \fett{v} \cdot \nabx \fett{v} = \frac e m \nabx \varphi\,, \qquad \nabx^2 \varphi = 4\pi (\rho - \bar\rho)\, , 
   \qquad \frac{\partial}{\partial t} \rho  + \nabx \cdot (\rho \fett{v}) =0 \,,
\ee
where now all fields depend on the Eulerian coordinate $\fett{x}$ and time $t$. See e.g.~\cite{Dawson,tutorialPlasma} for the application of perturbation techniques to equations~\eqref{eq:OCPeuler}.

\begin{figure}
  \begin{center}
  
   \includegraphics[width=0.9\textwidth]{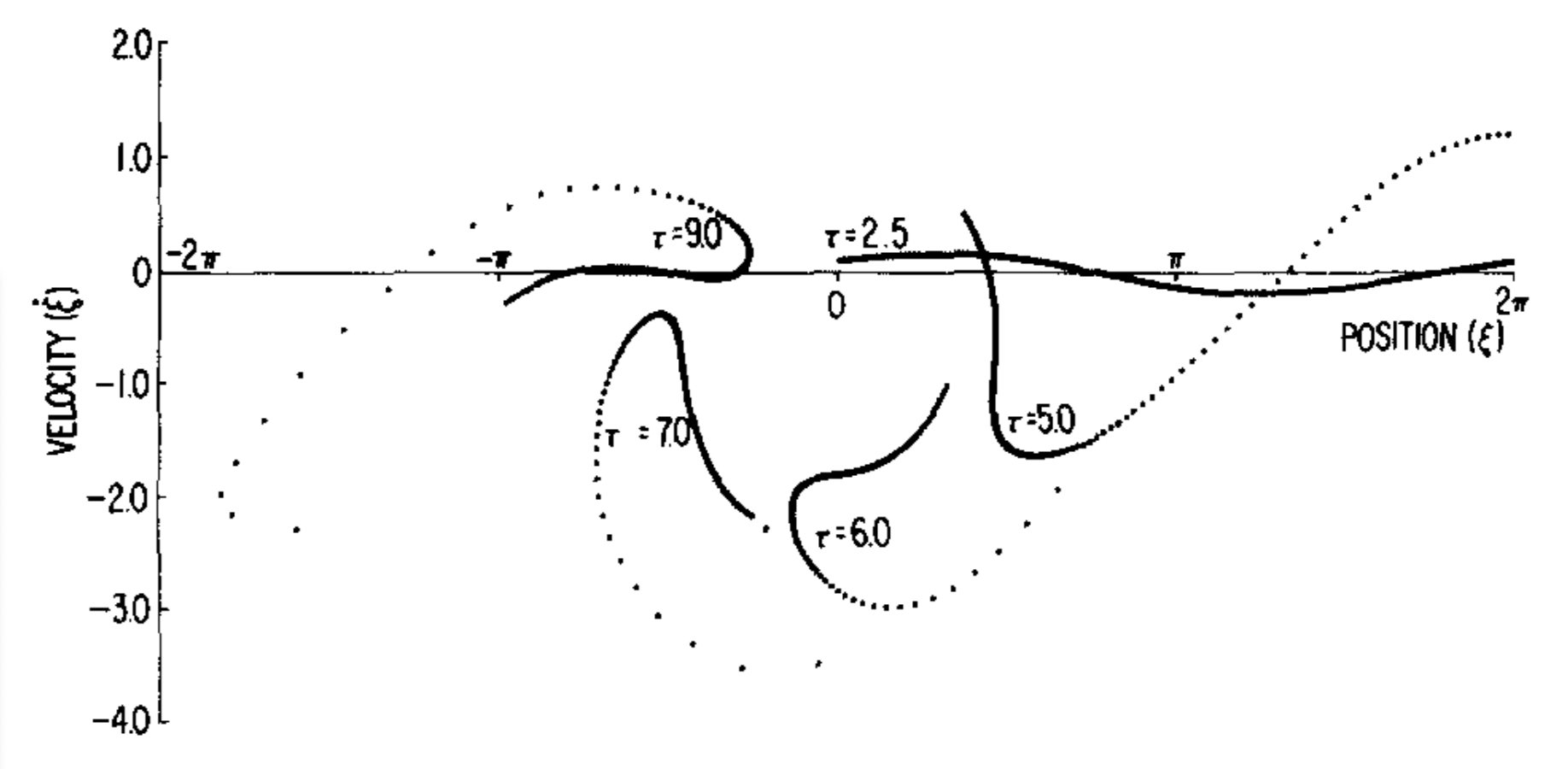}

  \end{center}
\caption{Phase-space for the beam-plasma instability, here in suitably rescaled coordinates. Figure from \cite{ONeil1971}; note that the filamentary tails of the distribution have been cut off for reasons of better visibility. }
\label{fig:plasma}
\end{figure}

\paragraph{Beam-plasma (two-stream) instability.}
A classical problem in plasma physics is when a beam of low-density, mono-energetic electrons is exposed to a cold thermalised, large-density plasma~\cite{Escande2018,ONeil1971,LesurDiamon2013}.
In the so-called single-wave model thereof, certain assumptions about the involved velocities are employed such that the plasma response is effectively non-resonant; therefore, the plasma can be treated as a bulk with respect to the propagation of the electron beam, while the plasma response can be incorporated in a form of a real dielectric function (see e.g.~\cite{Carlevaro2013,Carlevaro2020}).
Similarly as in the cosmological case, also here the Vlasov--Poisson equations are directly related to a Hamiltonian principle \cite{TennysonHamiltonianBOTI,Antoniazzi1HamiltonianBOTI}, and thus can be reduced to Newtonian-type equation of motion coupled to a Poisson equation. 
This set of equations has been provided by~\cite{ONeil1971} in the case of $N$~discretised  %(sheets of)
electrons in 1D, which, adopting the same notation as above, 
can be written as
\be \label{eq:eomP}
  \ddot x_{j}(t) =  \frac e m \nabla_x \varphi \,, \qquad \quad \nabla_x^2 \varphi = + 4\pi \rho_{\rm p}(x,t) + 4\pi \rho_{\rm b}(x,t) \,,
\ee
where $x_{j}$ denotes the current position of the $j$th electron, 
while $\rho_{\rm p}$ and $\rho_{\rm b}$ are respectively the charge densities of the plasma and the beam. 
It is usually assumed that  $\alpha := \rho_{\rm b}/\rho_{\rm p} \ll 1$ is a perturbatively small control parameter of the kinetic wave-particle interaction. Furthermore, it is assumed that the electron beam frequency is (roughly) equal to the plasma frequency, which ensures that the dielectric of the plasma is nearly vanishing \cite{Carlevaro2020}, thereby justifying  linear temporal expansions applied to the dielectric. This allows to incorporate the effect of the plasma density in the Poisson equation by a linear dielectric, and, at the same time, cast the Poisson equation into a simple evolution equation \cite{ONeilMalmberg}.
By contrast, the density of the electric beam is just a superposition 
of the $N$ charged particles, % i.e., 
\be \label{eq:charge}
 \rho_{\rm b}(x,t) = e \sum_{j=1}^N \delta_{\rm D}(x- x_j(t)) \,,
\ee
(see e.g. \cite{TennysonHamiltonianBOTI}).
Thus, the equations of motion for the beam-plasma instability are formally identical with the one of the OCP, equations~\eqref{eq:OCP}, however except with the addition that in the present case the Poisson equation receives a space and time-dependent dielectric function (see e.g.\ equation~14 in \cite{ONeil1971}).
We show in Fig.~\ref{fig:plasma} the corresponding phase-space of the beam–plasma instability:
At early times, the wave amplitude changes exponentially with the linear growth rate; this is thus a regime which can be determined using standard perturbative techniques \cite{ONeil1971}. At later times, the electrons are getting trapped, which in the figure is exemplified by the appearance of multi-beam regions that are sloshed back and forth.

Similarly as in the OCP case, one may employ a continuum description for investigating the beam-plasma instability. For this one can introduce the Lagrangian maps for the beam and plasma with $q \mapsto x^\alpha(q,t)$, where $\alpha =$p,b (see also \cite{FirpoElskens1998,Antoni1998} for alternative paths to a continuum description). The beam and plasma velocities are then defined with $v_\alpha(q,t) = \dot x_\alpha(q,t)$, and the Vlasov--Poisson equations are 
\be \label{eq:TSI}
  \ddot x_\alpha = \frac{e}{m}  (\nabla_x \varphi )_\alpha \,, \qquad \quad (\nabla_x^2 \varphi)_\alpha = + 4\pi \rho_{\rm p}(x_\alpha(q,t)) + 4\pi \rho_{\rm b}(x_\alpha(q,t)) 
\ee
for the components $\alpha =$p,b, where we have defined $(\nabla_x \varphi )_\alpha = \nabla_{x_\alpha} \varphi( x_\alpha(q,t))$, and mass conservation implies $\rho_{\rm b}(x_\alpha(q,t)) = \int \delta_{\rm D}(x_\alpha(q,t)- x_{\rm b}(q',t) )\dd q'$, and similarly for the plasma. Interestingly, equations~\eqref{eq:TSI} have a direct cosmological counterpart (in 3D) where they resemble the gravitationally coupled Vlasov--Poisson system of dark matter and visible matter/baryons \cite{Rampf:2020ety,Chen:2019cfu} (but only in the large-scale and late-time limit where baryons are effectively collisionless and pressureless).

The  mathematical methodology could also be extended to three space dimensions (see e.g. \cite{Escande2018}), as well as to the case of multiple cold beams, by e.g.\ adopting the formalism laid out in \cite{CarlevaroWarm}. 
Specifically, for $M$ cold beams, one essentially needs $M$ equations of motion~\eqref{eq:eomP} in the continuum limit coupled to a Poisson equation, where the latter then involves the superposition of $M$ charge densities~\eqref{eq:charge}.

\paragraph{The weak warm beam instability.}
Finally, another related plasma problem concerns the wave-particle description of Langmuir waves \cite{Escande2018,Carlevaro2013,BesseWarm}, where the latter is the response of a thermalised plasma when an electron beam is injected.   
The corresponding description is achieved when splitting the charged particles into so-called bulk and tail parts, which are subsequently dealt with differently. 
Here, the tail [bulk] part is the set of particles that is [not] resonant with Langmuir waves.
The beam appears as a bump on the tail of the overall velocity distribution function, and the so-called bump-on tail instability is excited through a region near the bump where the velocity distribution has positive slope.
 The tail particles are dealt with similarly as above, involving a source term as in the Poisson equation~\eqref{eq:charge}. By contrast, for the bulk particles, a quasi-ballistic approximation of particle trajectories is employed \cite{Escande2018,Antoni1998,ElskensPardoux}, which is obtained by setting for the trajectories $x_j = q_j + v_j t + \delta x_j^{\rm ini} + \delta \dot x_j^{\rm ini} t$  in our notation, where $\delta x_j^{\rm ini}$ and $\delta \dot x_j^{\rm ini}$ are the initial mismatches in position and velocities within the multi-beam-multi-array (a collection of monokinetic beams that provides a multi-valued velocity distribution). 
Observe here the structural similarities of this ballistic approximation compared against the Zel'dovich solution~\eqref{eq:ZA} or its higher-order generalisation~\eqref{eqs:LPT} in the cosmological case. We deem it possible to investigate the nature of the singularities (such as in the dielectric function of the plasma), possibly also involving methods adopted from catastrophe theory; see section~\ref{sec:strategy} and specifically section~\ref{sec:PSC}.

%%%%%%%%%%%%%%%%%%%%%%%%%%%%%%%
\section{Summary and outlook}\label{sec:conclusion}

Numerics should not be more than a few steps ahead of theoretical approaches. While this statement can be easily justified, it can be quite a challenge to accommodate in practice. 
This is particularly the case for Vlasov--Poisson descriptions that, due to the absence of collisions, have the strong tendency to form instabilities and singularities.

Solution methods for the cosmological  Vlasov--Poisson equations
are typically reserved for exploiting physically distinct regimes: theoretical calculations for the very early times of the gravitational evolution -- and brute-force numerical simulations that shed light into the highly non-linear regime of cosmic structures. Recently, however, the gap between theoretical and numerical methods  decreased  by exploiting at least three complementary avenues.

The first to mention are perturbative techniques, either by exploiting low-order truncations especially in Lagrangian space (section~\ref{sec:pertZA++}), and/or by using effective methods to Vlasov--Poisson (section~\ref{sec:effective}) that circumvent some of the obstacles, albeit in a rather pragmatic way. Still, these methods turn out to be extremely useful when interpreting data from current and forthcoming cosmological surveys.

The second avenue is novel simulation methods (section~\ref{sec:num}), that 
uncover much more details of the dark-matter phase-space as compared to standard particle-in-cell/$N$-body simulations. 
One of the simulation methods is solving for the tangent space of tracer particles (governed by a geodesic equation). 
Other methods employ phase-space tessellation techniques to obtain the dark-matter sheet, either constructed from an $N$-particle distribution, or by following the evolution of the vertices of simplices~(tetrahedra).

The third avenue addresses directly the mathematical skeleton of Vlasov--Poisson, which is made possible by exploiting the time-analyticity of dark-matter trajectories at sufficiently early times (section~\ref{sec:sc-random}). Straightforward (perturbative) expansion  techniques lead to the first non-trivial shell-crossing solutions, which are the phase-space locations when the number of dark-matter streams (beams) changes due to gravitational interactions. These expansion techniques are employed in Lagrangian coordinates, and allow one to analyse the dark-matter phase-space in a controlled set-up at spatio-temporal locations when the density is infinite (see Figs.\,\ref{fig:map-density}, \ref{fig:Saga} and \ref{fig:convergenceLPT}).

In general, analyticity in the dark-matter phase-space is lost directly at the first shell-crossing, due to the appearance of non-differentiable features in the particle accelerations (panels c and d in Fig.\,\ref{fig:PSC}).
However, thanks to a combination of an iteration technique (in the spirit of Picard's counterpart) as well as adapting Arnold's catastrophe theory, particle trajectories can be followed through shell-crossings by elementary means.

These theoretical methods are very accurate for times shortly after the first shell-crossing, essentially agreeing with independently performed high-resolution simulations. This agreement degrades at later times (Fig.\,\ref{fig:trajectories}), which could be rectified by including higher-order refinements. Nonetheless,
resolving the phase-space in dark-matter halos at late times, by elementary methods, remains a major challenge. 
 
Future directions on the cosmological side include the pairing of theoretical methods with simulations or machine-learning techniques. Also, in the longterm, the gained theoretical understanding will necessarily lead to improved (numerical) predictions  for cosmic structure formation.

With regard to the plasma case, we have laid out several intriguing connections to the cosmological case of Vlasov--Poisson. In particular, one may apply the continuous methods to the beam-plasma instability in the cold limit. Applications to the (warm) case of multi-beams are in principle straightforward, with the potential in contributing to the fundamental understanding of instabilities and chaos.

In the past, there has been a continuous transfer of knowledge between plasma physics and cosmology. In particular, many  numerical simulation techniques in cosmology have their roots in plasma physics. It is in the interest of scientific advancement that the multi-disciplinary transfer of knowledge thrives now and in the future.

\begin{acknowledgements}
We thank Patrick Diamond, Uriel Frisch, Oliver Hahn and Cora Uhlemann for many useful discussions and/or comments on the manuscript, as well as Mitsuru Kikuchi for the suggestion of writing this review.
\end{acknowledgements}

% Authors must disclose all relationships or interests that 
% could have direct or potential influence or impart bias on 
% the work: 
%
\begin{small}
\noindent{\bf Conflict of interest} The corresponding author declares that there is no conflict of interest.
\end{small}

%%%%%%%%%%%%%%%%%%%%%%%%%%

\end{document}